\newif\ifPosting\Postingtrue 

\ifPosting
\documentclass[sigplan,screen,authorversion]{acmart}
\startPage{1}
\else
\documentclass[sigplan,screen]{acmart}
\fi

\usepackage{flushend} 

\usepackage[nographicx,nonatbib,noxcolor,nohyperref,notheoremenvs,nosubfig]{bec}
\setboolean{showcomments}{false}
\usepackage[utf8]{inputenc}

\usepackage{booktabs}   
\usepackage{subcaption} 

\usepackage{stmaryrd}
\usepackage{dsfont}
\usepackage{mathtools,amsmath,amssymb}
\usepackage{proof}
\usepackage{fancyvrb}

\usepackage{tikz}
\usetikzlibrary{shapes, shapes.misc, calc, positioning, arrows}

\usepackage[nosfdefault]{comicneue}
\usepackage{fontawesome}

\newcommand{\comments}[1]{}

\newcommand{\revise}[1]{#1} 

\newcommand{\jedi}[1]{
 {\color{blue} \em #1}
}

\newcommand{\subst}[2]{[#1/#2]}
\newcommand{\wild}{*}
\newcommand{\mmid}{~\mid~}

\newcommand{\rname}[1]{\ensuremath{\textsf{#1}}}
\newcommand{\rn}[1]{\mbox{\scriptsize\textsf{#1}}}
\def\alignRuleX#1#2{\hbox{\raise -#1\baselineskip\hbox{#2}}}
\newcommand{\Rule}[3][]
 {\alignRuleX{0.50}{\infer[\!\rn{#1}]{#3}{#2}}}

\newcommand{\args}{\mathit{args}}

\newcommand{\vconcid}{\mathit{id}}
\newcommand{\vid}{\mathit{ID}}
\newcommand{\loc}{\mathit{XY}}
\newcommand{\tr}{\tau}
\newcommand{\trout}{\tr^{\text{\rm o}}}
\newcommand{\testop}{p}
\newcommand{\gen}{{\mathcal G}}

\newcommand{\gdom}[1]{\text{Gen}[#1]}

\newcommand{\prop}{{\mathcal P}}

\newcommand{\pbang}{!}

\newcommand{\aseq}{~\texttt{:>>}~}
\newcommand{\iseq}{~\texttt{*>>}~}
\newcommand{\iseqn}[1]{~\texttt{*>>}^{#1}~}
\newcommand{\choice}{~\texttt{<+>}~}

\newcommand{\bigaseq}{\texttt{:>>}}

\newcommand{\trstop}{\bullet}

\newcommand{\try}{~?}

\newcommand{\strD}{\mathds{S}}
\newcommand{\intD}{\mathds{Z}}

\newcommand{\tres}{\omega}

\newcommand{\tmenvE}[2]{\langle#1;#2\rangle}

\newcommand{\ttransstep}[3]{ #1 \rightarrowtail_{#2} #3 } 
\newcommand{\ttransstar}[3]{ #1 \rightarrowtail^{*}_{#2} #3 }

\newcommand{\devst}{\Phi}
\newcommand{\crrp}{{\mathcal R}}
\newcommand{\idleOrac}[1]{ \vdash_{\text{idle}} #1 }

\newcommand{\enbOrac}[3]{ #1 \vdash_{\text{enabled}} #2 \leadsto #3 }
\newcommand{\disabOrac}[2]{ #1 \vdash_{\text{disabled}} #2 }
\newcommand{\notenbOrac}[3]{ #1 \not \vdash_{\text{enabled}} #2 \leadsto #3 }
\newcommand{\notdisabOrac}[2]{ #1 \not \vdash_{\text{disabled}} #2 }
\newcommand{\crashOrac}[2]{ #1 \leadsto_{\text{crash}} #2 }
\newcommand{\mutOrac}[2]{ #1 \leadsto_{\text{mutate}} #2 }
\newcommand{\propOrac}[2]{ #1 \models_{\text{prop}} #2 }
\newcommand{\notpropOrac}[2]{ #1 \not \models_{\text{prop}} #2 }

\lstdefinelanguage{ChimpCheck}{
  morekeywords={Click,Swipe,Type,Rotate,ClickHome,ClickBack,Sleep,LongClick,Down,assert,optional,then},
  otherkeywords={:>>,*>>,<+>,==>,!,/\\},
  morekeywords=[2]{isDisplayed,isClickable,hasText},
  morekeywords=[3]{forAll,Gen,choose},
  morekeywords=[4]{val},
  morecomment=[l]{//},
  morecomment=[n]{/*}{*/},
  morestring=[b]",
  morestring=[b]',
  morestring=[b]"""
}
\newcommand{\codecc}[1]{\code[ChimpCheck]{#1}}

\makeatletter\if@ACM@journal\makeatother
\else\makeatother
\setcopyright{acmcopyright}
\acmPrice{15.00}
\acmDOI{10.1145/3133850.3133853}
\acmYear{2017}
\copyrightyear{2017}
\acmISBN{978-1-4503-5530-8/17/10}
\acmConference[Onward!'17]{2017 ACM SIGPLAN International Symposium on New Ideas, New Paradigms, and Reflections on Programming and Software}{October 25--27, 2017}{Vancouver, Canada}
\fi

\begin{document}

\title[Property-Based Randomized Test Generation for Interactive Apps]{ChimpCheck: Property-Based Randomized Test Generation for Interactive Apps} 
                                        
                                        


\author{Edmund S.L. Lam}
\affiliation{
  \institution{University of Colorado Boulder, USA}            
}
\email{edmund.lam@colorado.edu}          

\author{Peilun Zhang}
\affiliation{
  \institution{University of Colorado Boulder, USA}
}
\email{peilun.zhang@colorado.edu}

\author{Bor-Yuh Evan Chang}
\orcid{0000-0002-1954-0774}             
\affiliation{
  \institution{University of Colorado Boulder, USA}           
}
\email{evan.chang@colorado.edu}         


%
\begin{abstract}
We consider the problem of generating \emph{relevant} execution traces to test
rich interactive applications. Rich interactive applications, such as apps on
mobile platforms, are complex stateful and often distributed systems where
sufficiently exercising the app with user-interaction (UI) event sequences to
expose defects is both hard and time-consuming. In particular, there is a
fundamental tension between brute-force random UI exercising tools, which are
fully-automated but offer low relevance, and UI test scripts, which are manual
but offer high relevance. In this paper, we
\revise{consider a middle way---enabling a seamless fusion of scripted and
randomized UI testing. This fusion is prototyped in a testing tool called
ChimpCheck}
for programming, generating, and executing property-based randomized test cases
for Android apps. Our approach \revise{realizes this fusion} by offering a high-level,
embedded domain-specific language for defining custom generators of simulated
user-interaction event sequences. What follows is a combinator library built
on industrial strength frameworks for property-based testing (ScalaCheck) and
Android testing (Android JUnit and Espresso) to implement property-based
randomized testing for Android development. Driven by real, reported issues in
open source Android apps, we show, through case studies, how ChimpCheck enables
expressing effective testing patterns in a compact manner.
\end{abstract}







\maketitle

\newcommand{\mysection}[1]{\section{#1}}
\newcommand{\mysubsection}[1]{\subsection{#1}}
\newcommand{\myparagraph}[1]{\paragraph{#1}}

\mysection{Introduction} \label{sec:intro}

\comments{
The Android framework is essentially an event-driven system that brokers asynchronous 
interaction between internal framework events and user interaction events (UI events). 
Furthermore, high-expertise and experience is typically required 
to make full use of advance testing facilities of 
the Android testing framework (e.g., accessing active view hierarchy).
However, manually developing the necessary customized test cases one-at-a-time is extremely 
tedious and boiler-plate.
}

Driving Android apps to exercise relevant behavior is hard.
Android apps are complex stateful systems that interact with not only the vast Android framework but also a rich environment ranging from sensors to cloud-based services.
Even given a mock environment, the app developer must drive her app with a sufficiently large suite of user-interaction event sequences (UI traces) to test its ability to handle the multitude of asynchronous events dictated by the user (e.g., button clicks but also screen rotations and app suspensions).


Industrial practice of Android app testing is largely centered around two main techniques:
automated test generation via brute-force random UI exercising (Android 
Monkey~\cite{android-monkey}) and lower-level scripting of UI test cases 
(Android Espresso~\cite{android-espresso}, Robotium~\cite{robotium}). Monkey testing 
has the benefit of requiring very little development effort and offers a low-effort means 
to discover bugs in easily accessible regions of an app (that are environment independent). 
\comments{However, 
for better coverage (in the presence of environment-dependent code, e.g.,
getting past login pages) and 
testing of app-specific properties (e.g., checking the state of a media-player)
to exercise the app with specific, directed 
UI traces, to reach states of the app where specific properties can be checked.}%
\revise{However, to achieve more relevant testing,} the developer often has to rely on writing customized UI test scripts.
\revise{By {\em relevance}, we mean using application-specific knowledge to be exercise the app-under-test in a more sensible way.
For instance, to test a music-streaming service app, trying one failed login attempt is almost certainly sufficient.
Then, to exercise the interesting part of the app requires using a test account to get past the login screen to check, for example, that the media-player behaves in a way that users expect for a music service.}
While rich library support (e.g., Android JUnit, Espresso, and Robotium) and IDE integration (Android 
Studio) can make custom UI scripting more manageable, implementing test cases one-at-a-time to cover
all corner cases of an app is still a tedious and boilerplate process.

In the literature, 
\revise{advanced approaches for automating test generation has gained significant interest:
model-based techniques (e.g., Android Ripper~\cite{DBLP:journals/software/AmalfitanoFTTM15,DBLP:conf/kbse/AmalfitanoFTCM12}, Dynodroid~\cite{DBLP:conf/sigsoft/MachiryTN13},
evolutionary testing techniques (e.g., Evodroid~\cite{DBLP:conf/sigsoft/MahmoodMM14}) and search-based techniques 
(e.g., Sapienz~\cite{DBLP:conf/issta/MaoHJ16})}. While each of these techniques easily out performs 
pure random techniques, \revise{the development of all of these techniques have 
almost been entirely focused on pure automation. Little attention has been 
given to developing techniques that simplifies programmability and allowing higher-levels of customizability that
empowers the test developer to inject her app-specific knowledge into the test generation technique.
As a result, while these techniques offer effective automated solutions for 
testing generic functionality, 
they are unlikely to replace manual UI scripting because of their omission of the test developer's human insight and
app-specific knowledge.
Pure automation is unlikely to create generators, for example,
imbued with
knowledge to get pass a dynamic single-sign-on page.}


\comments{
Existing works on programmable interfaces for property-based randomized test generations~\cite{} 
are confined to generation of simple types and values, and algebraic datatypes. Hence,
the challenge of programming, generating and executing UI traces for simulating events 
in event-driven systems like Android, is yet unexplored and even largely undefined. 
Hence, commercial Android testing to date still heavily relies on boiler-plate UI test scripting 
to capture corner cases and domain-specific testing properties, but only uses test case generation 
by brute-force random or black-box model learning techniques to supplement the testing efforts.
}


\revise{\myparagraph{A New Paradigm for UI Testing}
The premise of this paper is that we cannot forsake human insight and app-specific knowledge. Instead, we must fuse scripted and randomized UI testing to derive \emph{relevant} test-case generators.
While improving and refining automated test generation techniques is indeed
a fruitful endeavor, an equally important research thread is developing expressive ways to
integrate human knowledge into these techniques.

As an initial demonstration of this fused approach, we present ChimpCheck, a proof-of-concept testing tool for programming, 
generating, and executing property-based randomized test cases for Android apps\footnote{ChimpCheck is available at~\url{https://github.com/cuplv/ChimpCheck}.}.}
Our key insight is that this tension between less relevant but automated and more relevant but manual can be eased or perhaps even eliminated by lifting the level of abstraction available to UI scripting.
Specifically, we make \emph{generators} of UI traces available to UI scripting, and we then discover that a brute-force random tester can simply be expressed as a particular generator. 
From a technical standpoint, ChimpCheck introduces property-based test case generation~\cite{DBLP:conf/icfp/ClaessenH00}
to Android by making user-interaction event sequences (i.e., UI traces) first-class objects,
\revise{enabling test developers to express (1) properties of UI traces that they deem to be relevant 
and (2) app-specific properties that are relevant to these UI traces.}
From a test developer's perspective,
\revise{ChimpCheck provides a high-level programming abstraction for writing UI test scripts 
and deriving app-specific test generators by integrating with advance test generation techniques---all from a single and simple programming interface.
Furthermore, regardless of the underlying generation techniques used, this
integrated framework generates a unified representation of relevant test artifacts (UI traces and generators), which effectively serves as both executable and human-readable specifications.}
%
To summarize, we make the following contributions:
\begin{itemize}
  \item We formalize a core language of user-interaction event sequences or
  \emph{UI traces} (Section~\ref{sec:synsem}). This core language captures what
  must be realized in a platform-specific test runner.  In ChimpCheck, the
  execution of UI traces is realized by the ChimpDriver component built on top
  of Android JUnit and Espresso. This formalization provides the foundation for
  generalizing property-based randomized test generation for interactive apps to
  other user-interactive platforms (e.g., iOS, web apps).
  \item Building on the formal notion of UI traces, we define \emph{UI trace
  generators} that lifts scripting user-interaction event sequences to scripting
  sets of sequences---po\-ten\-tial\-ly infinite sets of infinite sequences, conceptually (Section~\ref{sec:trace-generators}). This lifting enables the seamless mix of scripted and randomized UI testing. It also captures the platform-independent portion of ChimpCheck that compiles to the core UI trace language. This component is realized by building on top of ScalaCheck, which provides a means of sampling from generators.
  \item Driven by case studies from real Android apps and real reported issues, we demonstrate how
  ChimpCheck enables expressing customized testing patterns in a compact manner
  that direct randomized testing to produce relevant traces and targets specific
  kinds of bugs (Section~\ref{sec:chimp-in-action}). Concretely, we show testing
  patterns like interruptible sequencing, property preservation, \revise{randomized
  UI exercising}, and \revise{hybrid approaches} can all be expressed as derived
  generators, that is, generators expressed in terms of the core UI trace generators.
  %
\end{itemize}
\revise{In Section~\ref{sec:beyond}, we comment on how the ChimpCheck experience has motivated a vision for fused custom-scripting and automated-generation of interactive applications.}

\mysection{Overview: A Test Developer's View} \label{sec:overview}

In this section, we introduce \revise{our fused approach and ChimpCheck by means of an example testing scenario from the app-test developer's perspective. 
The purpose of this example is not necessarily to show every feature of ChimpCheck but rather to demonstrate the need to fuse app-specific knowledge with randomized testing.

\begin{figure}
  \includegraphics[width=\linewidth]{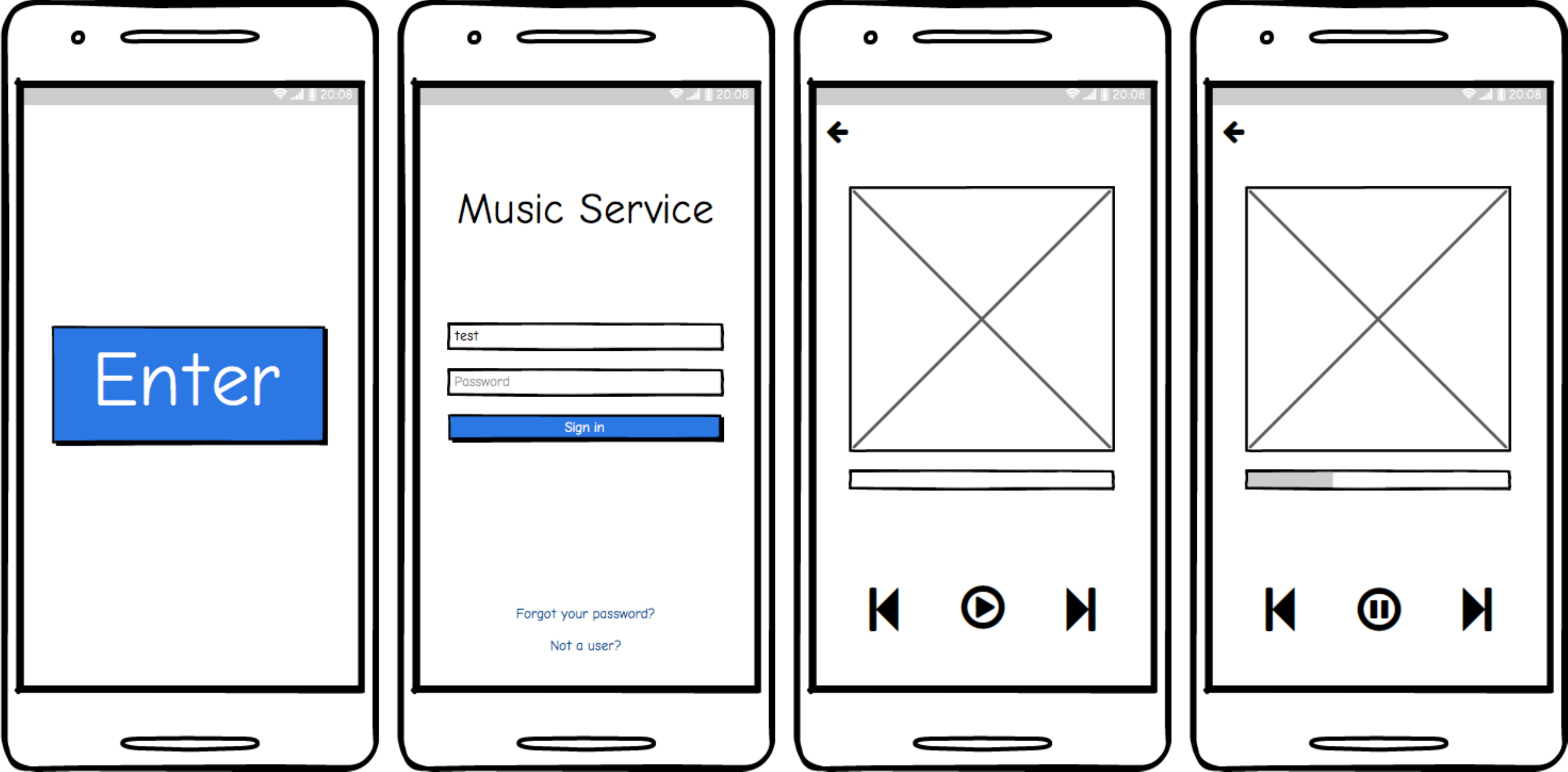}
\caption{\revise{Testing a music service app requires, for example, (1) fusing fixture-specific flows with random interrupts and (2) asserting app-specific properties.}}
\label{fig:music-service-app}
\end{figure}

Consider the problem of testing a music service app as shown in \figref{music-service-app}. Testing this app requires using test fixtures for getting past user authentication and asserting properties of the user interface specific to being a music player.
For concreteness, let us consider testing a particular user story for the app:
\begin{inparaenum}[(1)]
 \item\label{step-begin-fixture} Click on button {\comicneue Enter};
 \item Type in {\comicneue test} and {\comicneue 1234} into the username and password text boxes, respectively;
 \item\label{step-end-fixture} Click on button {\comicneue Sign in}; and
 \item\label{step-interesting-test} Click on buttons \faPlayCircleO{} and \faPauseCircleO{} to try out starting and stopping the music player, respectively.
\end{inparaenum}
Observe that the first few steps (Steps~\ref{step-begin-fixture}--\ref{step-end-fixture}) describe setting up a particular test fixture to get to the ``interesting part of the app,'' while the last step (\reftxt{Step}{step-interesting-test}) finally gets to testing the app component of interest.

This description captures the user story that the test developer seeks to validate, but an effective test suite will likely need more than one corresponding test case to see that that this user-flow through the app is robust.
With ChimpCheck, the test developer describes essentially the above script, but the script can be fused with fuzzing and properties to specify not a single test case but rather a family of test cases.
For example, suppose the test developer wants to generate a family of test cases where
\begin{enumerate}[A.]
  \item\label{overview-scenario-signin} The sign-in phase (Steps~\ref{step-begin-fixture}--\ref{step-end-fixture}) is robust to interrupt events such as screen rotations.
  \item The state of \faPlayCircleO{} and \faPauseCircleO{} buttons in the user interface corresponds in the expected manner to the internal state of an \code{android.media.MediaPlayer} object.
\end{enumerate}}%



\begin{figure}\small
\begin{lstlisting}[language=ChimpCheck,style=number,xleftmargin=20pt]
val signinTraces =#\label{line-signin-val-traces}#
 Click(R.id.enter) *>>#\label{line-signin-enter}#
 Type(R.id.username,"test") *>>#\label{line-signin-user}#
 { Type(R.id.password,"1234") *>>#\label{line-signin-passok}#
   Click(R.id.signin) *>>#\label{line-signin-signin}#
   assert(isDisplayed("Welcome")) } <+>#\label{line-signin-assertwelcome}\label{line-signin-or}#
 { Type(R.id.password,"bad") *>>#\label{line-signin-passbad}#
   Click(R.id.signin) *>>
   assert(isDisplayed("Invalid Password")) }#\label{line-signin-assertbad}#
   
forAll(signinTraces) { 
   trace => trace.chimpCheck() #\label{line-signin-checkit}#
}
\end{lstlisting}
\caption{ChimpCheck focuses test development to specifying the skeleton of user interactions---here, the valid and invalid sign-in flows for the music service app from \figref{music-service-app}.}
\label{fig:mediaplayer-eg1}
\end{figure}

\comments{
\jedi{
 In this section, provide 2 running motivating examples to show the following:
 \begin{enumerate}
  \item Why we can't just test vanilla sequences alone? (i.e., why the need of inserting interrupt events)
  \item Why property-based tests are necessary in Android?
 \end{enumerate}
 For 1, use scenario of a login page: where we need to execute sequence: (1) Click 'login', (2) 
 type 'test' in UserName box, (3) type '1234' in password box, (4) click 'go', (5) track 'welcome' view text is visible.
 Explain that we need to test this sequence with interrupt events inserted to track for either crash or unexpected
 behavior. E.g., after (3), screen rotate, then we cannot proceed to (4) because password box is empty. State that this
 is a very common bug in Android and cite some related bug reports where view elements vanish after screen rotate.   
 For 2, we further develop the example, state that just looking for app crash is often not enough: we often need to
 test for specific properties in an app. For instance, test if the app's text validator correctly handles valid/invalid inputs.
 Point out that 'valid' and 'invalid' are typically domain specific properties which black box testing cannot effectively
 test, hence certain amount of customization is always necessary. Setup: an app with a e-payment page. 
 Trace to test site: pass login, navigate to payment settings and test credit-card input with valid/invalid credit card number.
 State that failure in this scenario recently happened to uber: "unlimited free rides". 
}}

\myparagraph{A. Fusing Fixture-Specific Flows with Interrupts}

Testing requires sampling from valid and invalid scenarios.
\revise{While invalid sign-ins are easily sampled by brute-force random techniques, the most reasonable means of testing valid sign-ins is to hard-code a test fixture account.} While we would like a concise way of expressing such
fixtures, we also want some way to generate variations that test the sign-in process with interrupt events
(e.g., screen rotate, app suspend and resume) inserted at various points of the sign-in process. Such variations are important to test because
interrupts often are sources of crashes (e.g., null pointer exceptions) as well as unexpected behaviors (e.g., characters keyed into a text box vanishes after screen rotation). Developing such test cases one-at-a-time 
(via Espresso or Robotium~\cite{robotium,android-espresso}) is too time consuming, and most test-generation techniques
(via~\cite{DBLP:journals/software/AmalfitanoFTTM15,DBLP:conf/sigsoft/MachiryTN13,DBLP:conf/sigsoft/MahmoodMM14,DBLP:conf/issta/MaoHJ16}) would not be effective at finding valid sign-in sequences.


\revise{A key contribution of ChimpCheck is that it empowers the test developer to define the skeleton 
of the kind of UI traces of interest.
Concretely in \figref{mediaplayer-eg1}, we
specify not only a hard-coded sign-in sequence but variations of it that include interrupt actions that an
app-user could potentially interleave with the sign-in sequence.}

\revise{%
On \reftxt{line}{line-signin-val-traces}, the test developer defines a value \codecc{signinTraces} that is a \emph{generator} of sign-in traces where a trace is} a sequence of UI events that drives the app in some way.
\revise{To define \codecc{signinTraces}, the test developer describes essentially the sign-in flow outlined earlier:
\begin{inparaenum}[(1)]
 \item Click on button {\comicneue Enter} with \codecc{Click(R.id.enter)} on \reftxt{line}{line-signin-enter};
 \item Type in {\comicneue test}
 with \codecc{Type(R.id.username,"test")} on \reftxt{line}{line-signin-user}
 and type {\comicneue 1234}
 with \codecc{Type(R.id.password,"1234")} on \reftxt{line}{line-signin-passok};
 and
 \item Click on button {\comicneue Sign in}
 with \codecc{Click(R.id.signin)} on \reftxt{line}{line-signin-signin}.
\end{inparaenum}}
In Android, the \codecc{R} class contains generated constants that uniquely identify user-interface elements of an app. \revise{Like an Espresso-test developer today, we use these identifiers to name the user-interface elements.}

\revise{%
The user-interaction events like \codecc{Click} and \codecc{Type} specify an individual user action in a UI trace. These events can be composed together with a core operator like \codecc{:>>} that represents sequencing or,}
as on \reftxt{line}{line-signin-or},
with the operator \codecc{<+>} that implements a 
non-deterministic choice between its left operand (lines~\ref{line-signin-passok}--\ref{line-signin-signin})
and its right operand
(lines~\ref{line-signin-passbad}--\ref{line-signin-assertbad}).
\revise{%
Thus, \codecc{signinTraces} does not represent just a single trace but a set of traces. Here, we union two sets of traces to get some valid sign-ins (lines~\ref{line-signin-passok}--\ref{line-signin-assertwelcome}) and some invalid ones (lines~\ref{line-signin-passbad}--\ref{line-signin-assertbad}).
As the test developer, we can check for the
correctness of the sign-in scenarios with an \codecc{assert} for a welcome message in the case of the valid sign-ins (\reftxt{line}{line-signin-assertwelcome}) or for an ``Invalid Password'' error message in the case of the invalid ones (\reftxt{line}{line-signin-assertbad}). 
The \codecc{isDisplayed} expressions specify properties on the user interface that are then checked with \codecc{assert}.}

\revise{%
Finally, the UI events are sequenced together with the \codecc{*>>} combinator rather than the \codecc{:>>} operator.
The \codecc{*>>} combinator represents interruptible sequencing, that is, sequencing with some non-deterministic insertion of interrupt events.
Because these generator expressions represent sets of traces rather than just a single trace, this interruptible sequencing combinator \codecc{*>>} becomes a natural drop in for the core sequencing operator \codecc{:>>}. And indeed, interruptible sequencing \codecc{*>>} is definable in terms of sequencing \codecc{:>>} and other core operators.}


%
%

\begin{figure}\small
\begin{lstlisting}[language=ChimpCheck]
[1] Crashed after: Click(R.id.enter) :>> ClickHome
Stack trace: FATAL EXCEPTION: main, PID: 29302
java.lang.RuntimeException: Unable to start activity
...
[2] Failed assert isDisplayed("Welcome") after:
Click(R.id.enter) :>> Type(R.id.username,"test") :>>
Rotate :>> Type(R.id.password,"1234") :>>
Click(R.id.signin)
[3] Blocked after: Click(R.id.enter) :>> Rotate
\end{lstlisting}
\caption{Failed ChimpCheck test reports include the UI trace that leads to the crash or assertion failure.}
\label{fig:failed-login-reports}
\end{figure}

\revise{%
ChimpCheck is built on top of ScalaCheck~\cite{scalacheck}, extending this property-based test generation
library with UI traces and integration with Android test drivers running emulators or actual devices. From a practical standpoint, by building on top of ScalaCheck, ChimpCheck
inherits ScalaCheck's
test generation functionalities. The \texttt{signinTraces} generator is then passed to a \codecc{forAll} call on \reftxt{line}{line-signin-checkit}.
The \codecc{forAll} combinator comes directly from
the ScalaCheck library that generically implements sampling
from a generator. Tests are executed by invoking the ChimpCheck library operation \codecc{.chimpCheck()} on UI-trace samples. Here,
UI trace samples are bound to \codecc{trace} and executed on \reftxt{line}{line-signin-checkit} (i.e., \texttt{trace.chimpCheck()}).}
When this operation is executed, it triggers off a run of the app on an actual emulated Android device and submits
the trace \texttt{trace} (a trace sampled from \texttt{signinTraces}) to an associated test driver running in tandem with the app. When 
problems are encountered during each run, details of crashes or assertion failures are reported back to the user. 
\Figref{failed-login-reports} shows examples of bug reports from ChimpCheck.

An important contribution of ChimpCheck 
is that all reports contain the actual user-interaction trace executed up to the point of failure.
\revise{%
In an interactive application, a stack trace is severely limited because it contains only the internal method calls from the callback triggered by the last user interaction.}
This UI trace is essentially an executable
and concise description of the sequence of UI events that led up to the failure. And thus, these executed UI traces are valuable in the 
debugging process as they can guide the developer in reproducing the failure and ultimately understanding the root causes.

\myparagraph{B. Asserting App-Specific Properties}

%
\begin{figure}\small
\begin{lstlisting}[language=ChimpCheck,style=number,xleftmargin=0pt]
val playStateTraces =
 Click(R.id.enter) :>> Type(R.id.username,"test") :>>#\label{line-player-enter}#
 Type(R.id.password,"1234") :>> Click(R.id.signin) :>>#\label{line-player-signin}#
 optional {#\label{line-player-cycle-begin}#
  Click(R.id.play) *>> optional {
   Sleep(Gen.choose(0,5000)) *>> Click(R.id.stop)
  }
 }#\label{line-player-cycle-end}#
 
forAll(playStateTraces) { trace =>#\label{line-player-check-begin}#
 trace chimpCheck {
  (isClickable(R.id.play)#\,#==>#\,#!mediaPlayerIsPlaying) /\#\label{line-player-prop-begin}#
  (isClickable(R.id.stop)#\,#==>#\,#mediaPlayerIsPlaying)#\label{line-player-prop-end}#
 }
}#\label{line-player-check-end}#
\end{lstlisting}
\caption{ChimpCheck enables simultaneously describing the trace of relevant user interactions to drive to app to a particular state and checking properties on the resulting state. For the music service app from \figref{music-service-app}, we check that the state of the \faPlayCircleO{} (\codecc{R.id.play}) and \faPauseCircleO{} (\codecc{R.id.stop}) toggle is consistent with the state of the \codecc{MediaPlayer} object.}
\label{fig:mediaplayer-eg2}
\end{figure}

\revise{
Many problems in Android apps do not result in run-time exceptions that manifest as crashes but instead lead to unexpected behaviors.
It is thus critical that the testing framework provide explicit support for checking for app-specific properties.}
To check for unexpected behaviors,
the developer must write custom test scripts and invoke specific assertions at particular 
points of executing the Android app.
 
\revise{ChimpCheck provides explicit support for property-based testing, which
enables the test developer to simultaneously express generators for describing relevant UI traces and assertions for specifying app-specific properties to check.
In \figref{mediaplayer-eg2}, we show another UI trace generator \codecc{playStateTraces} for the same music-service app that focuses on testing that the UI state with the \faPlayCircleO{} and \faPauseCircleO{} toggle is consistent with the internal \code{MediaPlayer} object. Since the \codecc{signinTraces} from \figref{mediaplayer-eg1} already test the sign-in process, these tests simply wire-in the test fixture to get past the sign-in screen into the music player with the plain sequencing operator \codecc{:>>} (lines~\ref{line-player-enter}--\ref{line-player-signin}). 
Past the sign-in screen, lines~\ref{line-player-cycle-begin}--\ref{line-player-cycle-end} implement the optional-cycling between the \faPlayCircleO{} and \faPauseCircleO{} music player states via \codecc{Click}ing the corresponding buttons.
In this particular example, we insert a random idle of 0 to 5 seconds 
(\codecc{Sleep(Gen.choose(0,5000))}) between a cycle. Note that \codecc{Gen.choose(0,5000)} is not special to ChimpCheck; it is part of the ScalaCheck library to generate an integer between 0 and 5,000.}
Finally, on lines~\ref{line-player-check-begin}--\ref{line-player-check-end}, we 
execute the actual test runs as before, but this time we supply a property expression (lines~\ref{line-player-prop-begin}--\ref{line-player-prop-end}) comprising of two 
implication rules that asserts the expected consistency between the UI state and the underlying \code{MediaPlayer} state.
Note that while \codecc{isClickable} is a 
built-in atomic predicate that accesses the Android run-time view-hierarchy, \codecc{mediaPlayerIsPlaying} is a predicate defined by
the developer. Its implementation is simply a call to the \codecc{MediaPlayer} object's \codecc{isPlaying} method. 

\begin{figure*}[t]\centering\small
\(\begin{array}{c}
\text{(Strings)}~s\in\strD \quad \text{(Integers)}~n\in\intD \quad 
\text{(XY Coordinates)}~\loc ::= (n,n)
\quad \text{(UI Identifiers)}~\vconcid ::= n \mmid s \mmid \loc
\quad \vid ::= \vconcid \mmid \wild \\ \\[-1.5ex]

\text{(App Events)}~ a  
       ~::=~ \texttt{Click}(\vid) \mmid \texttt{LongClick}(\vid) \mmid \texttt{Swipe}(\vid,\loc) \mmid \texttt{Type}(\vid,s)
             \mmid \texttt{Pinch}(\loc,\loc) \mmid \texttt{Sleep}(n) \mmid \texttt{Skip} \\ \\[-1.5ex]
\text{(Device Events)}~ d
       ~::=~ \texttt{ClickBack} \mmid \texttt{ClickHome} \mmid \texttt{ClickMenu} \mmid \texttt{Settings} \mmid 
             \texttt{Rotate}
\qquad \text{(UI Events)}~u ~::=~ a \mmid d \\ \\[-1.5ex]

\text{(UI Traces)}~\tr ~::=~ u \mmid \tr_1 \aseq \tr_2 \mmid \texttt{assert}~\prop \mmid \tr\try 
                             \mmid \prop~\texttt{then}~\tr \mmid \trstop 
\qquad
\text{(Arguments)}~\args ~::=~ n \mmid s \mmid \args_1, \args_2 \\\\[-1.5ex]
\text{(Properties)} ~\prop ~::=~ \testop(\args) \mmid \pbang \prop \mmid \prop_1 \Rightarrow \prop_2 
                                 \mmid \prop_1 \wedge \prop_2 \mmid \prop_1 \vee \prop_2 

\end{array}
\)
\caption{A language of UI events, UI traces, and properties. UI events $u$ reify user interactions, and UI traces $\tr$ reify interaction sequences.}
\label{fig:traces_and_properties}
\end{figure*}

\mysection{User-Interaction Event Sequences} \label{sec:synsem}

This section defines a core language of UI traces to understand what must be realized in a platform-specific test runner. We first discuss the syntactic constructs that make
up this language (Section~\ref{ssec:ui-trace}). Then, we present an operational semantics that gives these declarative symbols formal meaning by defining the interactions between UI traces and an abstraction of
the device and app in test (Section~\ref{ssec:trace-sem}).
This reification of implicit user interactions into explicit UI traces is crucial for fusing scripting and randomized testing.
We finally discuss how an instance of these semantics are realized for testing Android apps (Section~\ref{ssec:uitrace-implementation}).
\comments{
We introduce the language and semantics of UI Traces and properties. Following these, we discuss various
engineering challenges which we have overcome to achieve an implementation of this semantics.
}

\mysubsection{A Language of UI Traces and Properties}
\label{ssec:ui-trace}

\comments{
\jedi{
Introduce the syntax of UI event traces and property assertions. Explain basics of the language (e.g., what are IDs and XYs, 
and that properties are just propositional logic formulas). Describe about the $*$ operator we call 'wild card' and that it is 
a high-level way for the developer to access the runtime view hierarchy.
Explain that implementation section will highlight how we achieve this.
Mention that property predicates $\testop(\args)$ maps to both built-in operations provided by Android testing framework
(e.g., isClickable(), isEnabled()), as well as user defined runtime test operations on the test driver.
}}

Figure~\ref{fig:traces_and_properties} shows the syntax of the language of UI traces
and properties. Basic terms of the language comprise of strings of characters ($s$) and
integers ($n$). Specific points ($\loc$ coordinates) of the device display that renders 
the app are expressed by a 2-tuple of integers.
UI elements can be identified ($\vid$) with an integer identifier $n$, the display text of the element $s$, or the element's $\loc$ coordinate.
Additionally, our language includes a {\em UI element wild card} \texttt{*}, which represents a reference to some existing UI element without committing to a specific element.
Existing testing frameworks 
for Android apps (e.g., Robotium and Espresso) enable developers to reference particular UI elements using 
integer identifiers (generated constants found in the \texttt{R.id} module of 
an Android app),
display text, and $\loc$ coordinates.
The UI element wild card \texttt{*} is the lowest level or simplest example of fusing scripting and test-case generation (see Section~\ref{ssec:uitrace-implementation} for further details).

App events ($a$) represent the UI events that a user can apply onto an app while staying within the app.
Device events ($d$) are
interrupt events that potentially suspends and resumes the app. UI traces $\tr$ are compositions
of such events with some richer combinators: our example earlier introduced
sequencing $\tr_1 \aseq \tr_2$ and assertions $\texttt{assert}~\prop$, informally the try
operator $\tr\try$ represents an attempt on $\tr$ that will not halt the test (unless the app
crashes or violates an assert), $\prop~\texttt{then}~\tr$ represents a trace $\tr$ guarded by a
property $\prop$, while $\trstop$ is the unit operator. The language of properties
is a standard propositional logic with the usual interpretation. 
Predicates \texttt{p} can be either user-defined predicates (e.g., \texttt{mediaPlayerIsPlaying} 
in Figure~\ref{fig:mediaplayer-eg2}) or built-in predicates that maps to known test operations
provided by the Android framework (e.g., \texttt{isClickable}, \texttt{isDisplayed}). We
discuss implementation of these predicates in Section~\ref{ssec:uitrace-implementation}.

What is critical here is not, for example, the particular app and device events but rather the reification of user interactions into UI traces $\tr$.

\mysubsection{A Semantics of UI Traces and Properties}
\label{ssec:trace-sem}

\begin{figure}\centering\small
\(\begin{array}{c}
  \text{(Device States)} \quad \devst \qquad \text{(Crash Reports)} \quad \crrp \\[1ex]
  \text{(Oracles)}
  \begin{array}{c}
    \quad \enbOrac{\devst}{u}{\devst'} \qquad \disabOrac{\devst}{u} \qquad \idleOrac{\devst} \\[0.5ex]
    \quad \propOrac{\devst}{\prop} \qquad \mutOrac{\devst}{\devst'} \qquad \crashOrac{\devst}{\crrp} 
  \end{array} \\[3ex]
  \text{(Results)} ~ \tres ~::=~ \texttt{succ} \mmid \texttt{crash}~\crrp \mmid \texttt{fail}~\prop \mmid \texttt{block}~u \\  
  \text{(Exec Events)} ~ o  ~::=~ u \mmid \trstop \quad
  \text{(Exec Traces)} ~ \trout  ~::=~  o \mmid \trout_1 \aseq \trout_2
\end{array}\)
\caption{Oracle judgments abstract application and device-specific transitions.}
\label{fig:device_state}
\end{figure}

\comments{
\jedi{
Introduce our abstraction of the Android framework, and the other auxiliary computation artifacts that
comprises of the state of our system, as shown by Figure~\ref{fig:device_state}. Introduce the term
'Chimp Driver', say that it is what we call the runtime test driver that interprets the chimpcheck
trace combinators. Discuss Figure~\ref{fig:trace_semantics}, the small-step and big-step operational 
semantics of Chimp Driver. Explain that the outputs of the transitions are the reproducible trace
$\trout$ and result $\tres$.
}}

\begin{figure*}\centering\small
\(\begin{array}{@{}c@{}}
  \begin{array}{@{}c}
     \fbox{\text{(Chimp Driver Small-Step Transitions)} 
           $\qquad \ttransstep{\tmenvE{\tr}{\devst}}{o}{\tmenvE{\tr'}{\devst'}} \qquad 
                   \ttransstep{\tmenvE{\tr}{\devst}}{o}{\tmenvE{\tres}{\devst'}}$} \\\\[-1.5ex]
     \Rule[(Mutate)]
          { \mutOrac{\devst}{\devst'} }
          { \ttransstep{\tmenvE{\tr}{\devst}}{\trstop}{\tmenvE{\tr}{\devst'}} }
     \qquad
     \Rule[(Crash)]
          { \crashOrac{\devst}{\crrp} }
          { \ttransstep{\tmenvE{\tr}{\devst}}{\trstop}{\tmenvE{\texttt{crash}~\crrp}{\devst}} } 
     \qquad
     \Rule[(No-Op)]
          {  }
          { \ttransstep{\tmenvE{\trstop}{\devst}}{\trstop}{\tmenvE{\texttt{succ}}{\devst}} }
     \\ \\[-1.5ex]
     \Rule[(Enabled)]
          { \enbOrac{\devst}{u}{\devst'} }
          { \ttransstep{\tmenvE{u}{\devst}}{u}{\tmenvE{\trstop}{\devst'}} } 
     \quad
     \Rule[(Block-1)]
          { \disabOrac{\devst}{u} }
          { \ttransstep{\tmenvE{u}{\devst}}{\trstop}{\tmenvE{\texttt{block}~u}{\devst}} }
     \quad
     \Rule[(Inferred)]
          { \enbOrac{\devst}{\subst{\vconcid}{\wild}a}{\devst'} }
          { \ttransstep{\tmenvE{a}{\devst}}{\subst{\vconcid}{\wild}a}{\tmenvE{\trstop}{\devst'}} } 
     \quad
     \Rule[(Block-2)]
          { \forall id \{ \disabOrac{\devst}{\subst{\vconcid}{\wild}a} \} }
          { \ttransstep{\tmenvE{a}{\devst}}{\trstop}{\tmenvE{\texttt{block}~a}{\devst}} }
     \\ \\[-1.5ex]
     \Rule[(Seq-1)]
          { \ttransstep{\tmenvE{\tr_1}{\devst}}{o}{\tmenvE{\tr_1'}{\devst'}}}
          { \ttransstep{\tmenvE{\tr_1 \aseq \tr_2 }{\devst}}{o}{\tmenvE{\tr_1' \aseq \tr_2}{\devst'}} }
     \qquad
     \Rule[(Seq-2)]
          { \idleOrac{\devst} }
          { \ttransstep{\tmenvE{\trstop \aseq \tr_2 }{\devst}}{\trstop}{\tmenvE{\tr_2}{\devst}} }
     \qquad
     \Rule[(Seq-3)]
          { \ttransstep{\tmenvE{\tr_1}{\devst}}{o}{\tmenvE{\tres}{\devst'}}
            \quad \tres \neq \texttt{succ} }
          { \ttransstep{\tmenvE{\tr_1 \aseq \tr_2}{\devst}}{o}{\tmenvE{\tres}{\devst'}} }
     \\ \\[-1.5ex]
     \Rule[(Assert-Pass)]
          { \propOrac{\devst}{\prop} }
          { \ttransstep{\tmenvE{\texttt{assert}~\prop}{\devst}}{\trstop}{\tmenvE{\trstop}{\devst}} } 
     \qquad
     \Rule[(Assert-Fail)]
          { \notpropOrac{\devst}{\prop} }
          { \ttransstep{\tmenvE{\texttt{assert}~\prop}{\devst}}{\trstop}{\tmenvE{\texttt{fail}~\prop}{\devst}} } 
     \\ \\[-1.5ex]
     \Rule[(Try-1)]
          { \ttransstep{\tmenvE{\tr}{\devst}}{o}{\tmenvE{\tr'}{\devst'}} }
          { \ttransstep{\tmenvE{\tr\try}{\devst}}{o}{\tmenvE{\tr'\try}{\devst'}} }
     \quad
     \Rule[(Try-2)]
          { \ttransstep{\tmenvE{\tr}{\devst}}{o}{\tmenvE{\tres}{\devst'}}  \quad \tres \!\in\! \{\texttt{succ},\texttt{block}~u\} }
          { \ttransstep{\tmenvE{\tr\try}{\devst}}{o}{\tmenvE{\trstop}{\devst'}} }
     \quad
     \Rule[(Try-3)]
          { \ttransstep{\tmenvE{\tr}{\devst}}{o}{\tmenvE{\tres}{\devst'}} \quad \tres \!\in\! \{\texttt{crash}~\crrp,\texttt{fail}~\prop\}}
          { \ttransstep{\tmenvE{\tr\try}{\devst}}{o}{\tmenvE{\tres}{\devst'}} }
     \\ \\[-1.5ex]
     \Rule[(Qualified)]
          { \propOrac{\devst}{\prop} }
          { \ttransstep{\tmenvE{\prop~\texttt{then}~\tr}{\devst}}{\trstop}{\tmenvE{\tr}{\devst}} }
     \qquad
     \Rule[(Unqualified)]
          { \notpropOrac{\devst}{\prop} }
          { \ttransstep{\tmenvE{\prop~\texttt{then}~\tr}{\devst}}{\trstop}{\tmenvE{\trstop}{\devst}} }
  \end{array} \\ \\[1ex]
  \begin{array}{@{}c@{}}
     \fbox{\text{(Chimp Driver Big-Step Transitions)} 
           $\qquad \ttransstar{\tmenvE{\tr}{\devst}}{\trout}{\tmenvE{\tres}{\devst'}}$} \\ \\[-1.5ex]
     \Rule[Trans]
          { \ttransstep{\tmenvE{\tr}{\devst}}{o}{\tmenvE{\tr'}{\devst'}}
            \quad
            \ttransstar{\tmenvE{\tr'}{\devst'}}{\trout}{\tmenvE{\tres}{\devst''}} }
          { \ttransstar{\tmenvE{\tr}{\devst}}{(o \aseq \trout)}{\tmenvE{\tres}{\devst''}} } 
     \qquad
     \Rule[End]
          { \ttransstep{\tmenvE{\tr}{\devst}}{o}{\tmenvE{\tres}{\devst'}} }
          { \ttransstar{\tmenvE{\tr}{\devst}}{o}{\tmenvE{\tres}{\devst'}} }
  \end{array} 
\end{array}\)
\caption{Chimp Driver is the operational semantics that defines the interpretation of UI traces $\tr$ in terms of the application and device-specific oracle judgments in Figure~\ref{fig:device_state}.} 
\label{fig:trace_semantics}
\end{figure*}

Our main focus in this subsection is the operational semantics of a transition system
we call Chimp Driver, that interprets UI events and submits commands to invoke the relevant UI action. That is, these semantics give an interpretation to reified user interactions.

Apps are complex, stateful event-driven systems. To abstract over the particulars of any particular app and its complex interactions with the underlying event-driven framework, we define Chimp Driver in terms of
several {\em oracle} judgments (shown in Figure~\ref{fig:device_state}).
These oracle judgments are then realized in an implementation on a per platform basis. In 
Section~\ref{ssec:uitrace-implementation},
we discuss how we realize these oracle judgments for Android apps.
The device state $\devst$ represents the consolidated state of the app and the (Android) device. 
The oracle judgment form $\enbOrac{\devst}{u}{\devst'}$ describes a transition of the device state from $\devst$ to $\devst'$ on seeing
UI event $u$.
Recall that UI events $u$ are primitive symbols of UI traces and they are ultimately
mapped to relevant actions on the UI (e.g., $\texttt{Click}(ID)$ to click on view with identifier $ID$).
An event $u$ is said to be {\em concrete} (denoted by $\mathit{concrete}(u)$) if and only if $u$ does not
contain an argument with the UI element wild card $\texttt{*}$ (i.e., all $\vid$s are $\vconcid$s). The {\em enabled} oracle is only defined for concrete events.
The judgment $\disabOrac{\devst}{u}$ holds if an event $u$ is not applicable in the current state $\devst$.
Similarly, it is only defined for concrete events. We assume that these two judgments are mutually exclusive, formally:
\[\begin{array}{@{}c@{}}
  \forall~ \devst,\devst',u~ (\mathit{concrete}(u) \wedge \enbOrac{\devst}{u}{\devst'} ~\Rightarrow~ \notdisabOrac{\devst}{u}) \\
  \forall~ \devst,\devst',u~ (\mathit{concrete}(u) \wedge \disabOrac{\devst}{u} ~\Rightarrow~ \notenbOrac{\devst}{u}{\devst'})
\end{array}\]
Note that $\enbOrac{\devst}{u}{\devst'}$ is possibly asynchronous and imposes no constraints that in fact
the action corresponding to $u$ has been executed. For this, we rely on the judgment 
$\idleOrac{\devst}$, that holds if the test app's main thread has executed all previous actions and
is idling. The judgment $\propOrac{\devst}{\prop}$ holds if a given property $\prop$ holds in the current
state $\devst$. Since $\prop$ is a fragment of propositional logic, its decidability depends
on the decidability of the interpretations of predicates $p$.
Since the language of properties is standard propositional logic, we omit detailed definitions.
The judgment $\mutOrac{\devst}{\devst'}$ describes a transition of the device from $\devst$ to $\devst'$
that is not a direct consequence of Chimp Driver operations. This corresponds to background asynchronous tasks
that can be either part of the test app or other running tasks of the device. 
Note that $\idleOrac{\devst}$ does not imply that $\mutOrac{\devst}{\devst'}$ is not possible but
simply that the state of the main thread appears idle.
This interpretation, of course, ultimately results in certain impreciseness in the reports extracted by Chimp Driver
(as certain race conditions are in-distinguishable), 
but such is a known and expected consequence of UI testing. Finally, $\crashOrac{\devst}{\crrp}$ observes
a transition of the device to a crash state, with a crash report $\crrp$ (conceptually, a stack trace from handling the event the ends in a crash).

The state of the Chimp Driver is the pair $\tmenvE{\tr}{\devst}$ comprising of the current UI trace $\tr$ and current
device state $\devst$. Results $\tres$ are terminal states of this transition system and come in
four forms: $\texttt{succ}$ indicates a successful run, $\texttt{crash}~\crrp$ indicates a crash with report $\crrp$,
$\texttt{fail}~\prop$ indicates a failure to assert $\prop$, and $\texttt{block}~u$ indicates that the Chimp Driver
got stuck while attempting to execute event $u$. An auxiliary output of Chimp Driver is the actual concrete trace that was
executed: $\trout$ is a sequencing (\texttt{:>>}) of primitive events $u$ or the nullipotent (unit) event $\trstop$. 
It is the executed UI trace that reproduces the failure (modulo race-conditions).

Figure~\ref{fig:trace_semantics} introduces the operational semantics of Chimp Driver. Its small-step semantics is 
defined by the transitions $\ttransstep{\tmenvE{\tr}{\devst}}{o}{\tmenvE{\tr'}{\devst'}}$ and 
$\ttransstep{\tmenvE{\tr}{\devst}}{o}{\tmenvE{\tres}{\devst'}}$ that defines intermediate and terminal transitions
respectively. Intuitively, $\tr$ and $\devst$ are inputs of the transitions, while $\tr'$, $\tres$, $\devst'$ together
with the executed event $o$, are the outputs. The big-step semantics is defined by the transition
$\ttransstar{\tmenvE{\tr}{\devst}}{\trout}{\tmenvE{\tres}{\devst'}}$ that exhaustively chains together steps of 
the small-step transitions. Outputs of the test run are $\trout$, $\tres$, and possibly observable (from the
confines of the device framework) fragments of $\devst'$.
The following paragraphs explains the purpose of each transition rule presented in Figure~\ref{fig:trace_semantics}.

\myparagraph{Mutate, Crash, and Unit}
The \rname{(Mutate)} and \rname{(Crash)} rules lift mutate and crash oracle 
transitions into Chimp Driver transitions. While the \rname{(Mutate)} transition is transparent to Chimp Driver,
\rname{(Crash)} results in a terminal $\texttt{crash}~\crrp$ state. \rname{(No-Op)} transits the unit event
$\trstop$ to a success state $\texttt{succ}$. 

\myparagraph{UI Events}
The four rules
$\rname{(Enabled)}$, $\rname{(Block-1)}$, $\rname{(Inferred)}$, and $\rname{(Block-2)}$ define the UI event transitions. The $\rname{(Enabled)}$
rule defines the case when a concrete event $u$ is successfully inserted into the device state, while $\rname{(Block-1)}$
defines the case when $u$ is not enabled and the execution blocks. Note that having $\texttt{block}~u$ as an output 
is an important diagnostic behavior of Chimp Driver, hence this blocking behavior is explicitly modeled in the
transition system. The next two rules only apply to non-concrete events (containing $\texttt{*}$ as argument):
the $\rname{(Inferred)}$ rule defines the case when a non-concrete event $a$ is successfully concretized,
during which the occurrence of $\wild$ in $a$ is substituted by some UI identifier $id$ (if it exists) such that the instance of $a$ is an enabled event. This substitution is denoted by $\subst{\vconcid}{\wild}a$.
Finally the rule $\rname{(Block-2)}$ defines the case when no enabled events can be inferred from instantiating
$\wild$ to any valid UI identifier (i.e., all valid $id$'s are disabled), hence the transition system blocks. 

\myparagraph{UI Event Sequencing}
UI event sequencing ($\tr_1 \aseq \tr_2$) is defined by three rules: the rule $(\texttt{Seq-1})$ defines a 
single-step transition on the prefix $\tr_1$ to $\tr_1'$. The rule $(\texttt{Seq-2})$ defines the case when
the prefix trace is a unit event $\trstop$, during which the derivation can only proceed if the device is
in an idle state (i.e., $\idleOrac{\devst}$). Finally the rule $(\texttt{Seq-3})$ defines the case when the prefix
trace ends in some failure result ($\tres \neq \texttt{succ}$), during which the transition system terminates with
$\tres$ as the final result. Importantly, note that the purpose of having $\idleOrac{\devst}$ as a premise of
the $\rname{(Seq-2)}$ rule: this condition essentially enforces the execution of Chimp 
Driver {\em in tandem} with the test app. Particularly, this means that any event $u$ in the prefix sequence $\tr_1$ must have been executed before postfix $\tr_2$ is attempted. We assume that the actual execution of these events are modeled by unspecified $\rname{(Mutate)}$ transitions and that the state of the device eventually reaches an idle
state (i.e., $\idleOrac{\devst}$) once all pending events have been processed.

This semantics supports an intuitive, synchronous view of UI traces.
For instance, consider the following UI trace:
\begin{lstlisting}[language=ChimpCheck]
  assert count(0) :>> Click(R.id.cnt) :>>
  assert count(1) :>> Click(R.id.cnt) :>>
  assert count(2) :>> $\ldots$
\end{lstlisting}
In this example, $\texttt{count}(n)$ is a predicate that asserts that the UI element $\texttt{R.id.cnt}$
has so far been clicked $n$ times. Notice that the correctness of this UI trace is predicated on the 
assumption that each $\texttt{Click}(R.id.cnt)$ operation is actually synchronous and in tandem with the 
test app. This is modeled by the premise $\idleOrac{\devst}$ of the \rname{(Seq-2)} rule, and we will discuss
its implementation in the Chimp Driver for Android in the next section.
The mainstream Android testing frameworks, such as Espresso and Robotium, have
gone through great lengths to achieve this synchronous interpretation and publicly cite this behavior as a benefit of the frameworks.

\myparagraph{Assertions and Try}

The $\rname{(Assert-Pass)}$ and $\rname{(Assert-Fail)}$ rules handle assertions. They each consult
the property oracles $\propOrac{\devst}{\prop}$ and $\notpropOrac{\devst}{\prop}$ to result
in a success ($\trstop$) or failure ($\texttt{fail}~\prop$), respectively.
The {\em try} combinator ($\tr\try$) represents an attempt to execute trace $\tr$ that should not terminate the existing test
(rule $\rname{(Try-2)}$) unless $\tr$ results in a crash or failed assertion (rule $\rname{(Try-3)}$).
Rule $\rname{(Try-1)}$ simply defines intermediate transitions. From the test developer perspective, she writes ($\tr\try$)
when she wants to suppress the blocking behavior within the trace $\tr$ and just move on to the rest of the UI sequence
(if any). We discuss the motivation for this combinator in Section~\ref{ssec:action-random}.

\myparagraph{Conditional Events}

The $\rname{(Qualified)}$ and $\rname{(Unqualified)}$ rul\-es handle cases of the combinator 
$\prop~\texttt{then}~\tr$. Informally, this combinator executes $\tr$ only if $\prop$ holds (i.e., 
$\propOrac{\devst}{\prop}$). This combinator represents the language's ability to express conditional interactions depending on the device's run-time state
that are often necessary when an app's navigational behavior is not static
(see Section~\ref{ssec:action-injection} for examples).

\mysubsection{Implementing Chimp Driver for Android}
\label{ssec:uitrace-implementation}

Here, we discuss one example instance of implementing Chimp Driver (along with the oracle judgments) abstractly defined in Section~\ref{ssec:trace-sem}.

\begin{figure}\centering\small
\begin{tikzpicture}[scale=0.9, every node/.style={scale=0.9}]

\node[draw=black,style={scale=1}] (Dev) at (3,2) { 
   \renewcommand{\arraystretch}{0.8}
   \begin{tabular}{c}
     Test \\ Developer 
   \end{tabular}
};

\node[draw=black,style={scale=1}] (Cmb) at (-0.5,0) { 
   \renewcommand{\arraystretch}{0.8}
   \begin{tabular}{c}
     Combinator \\ Library \\
     (Scala)
   \end{tabular}
};

\node[draw=black,style={scale=1}] (Coor) at (5,0) { 
   \renewcommand{\arraystretch}{0.8}
   \begin{tabular}{c}
     Test Controller \\
     $\left( 
      \begin{array}{c}
       \text{Scala},~\text{Google Protobuf}, \\
       \text{Android Adb}, \\
       \text{Instrumentation APIs}
      \end{array}
     \right)$
   \end{tabular}
};

\node[draw=black,style={scale=1}] (Drv) at (2,-2.5) { 
   \renewcommand{\arraystretch}{0.8}
   \begin{tabular}{c}
     Chimp Driver \\
     $\left( 
      \begin{array}{c}
       \text{Java},~\text{Android JUnit}, \\
       \text{Android Espresso}
      \end{array}
     \right)$ \\
     $\ttransstar{\tmenvE{\tr}{\devst}}{\trout}{\tmenvE{\tres}{\devst'}}$
   \end{tabular}
};

\draw[->,line width=1.2pt] (Dev.west) -| (Cmb.north) node[midway,above] {Develops test scripts} ;
\draw[->,line width=1.2pt] (Coor.north) |- (Dev.east) node[midway,right] {
   \begin{tabular}{c}
     Triages \\
     outcome \\
     $\trout$ and $\tres$
   \end{tabular}
} ;
\draw[->,line width=1.2pt] (Cmb.east)--(Coor.west) node[midway,above]{UI trace $\tr$};

\draw[-,line width=1.2pt] (Coor.south)+(-1.5,0) |- (3,-1.3);
\draw[->,line width=1.2pt] (3,-1.3) -| (Drv.north) node[midway,above]{Protobuf encoded $\tr \qquad \quad$};

\draw[->,line width=1.2pt] (Drv.east) -| (Coor.south) node[near end,right]{
   \begin{tabular}{c}
     Protobuf \\
     encoded \\
     $\trout$ and $\tres$
   \end{tabular}
};

\end{tikzpicture}
\caption{Implementing Chimp Driver for Android. Here, we depict the development and execution of a single UI trace $\tr$. The Test Controller runs on the testing server, and Chimp Driver runs on the device. Both are Android-specific implementations.}
\label{fig:archit-1}
\end{figure}
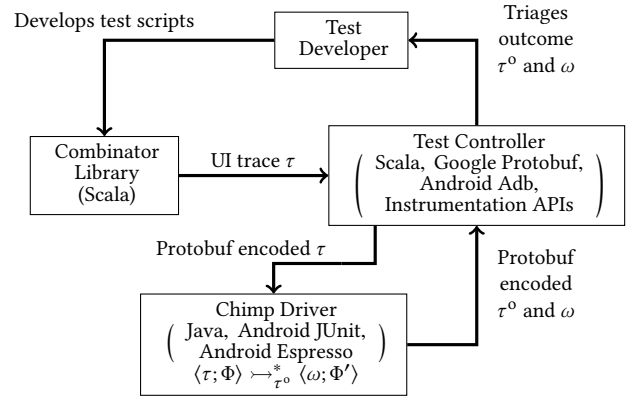

\myparagraph{ChimpCheck Run-time Architecture} 
Figure~\ref{fig:archit-1} illustrates the single-trace architecture of the ChimpCheck combinator
library and the test driver (Chimp Driver) for Android. Here, we depict the scenario where the test developer
programs single UI traces directly; in Section~\ref{ssec:generator-implement}, we describe how we lift this architecture to sets of traces.
The main motivation of our development work here
is to implement a system that can be readily integrated with existing Android development
practices.

The {\em combinator library} of ChimpCheck is implemented in Scala. Our reason for choosing
Scala is simple and well-justified. Firstly, Scala programs integrate well with Java
programs, hence importing dependencies from the underlying Android Java projects is
trivial. For instance, it would be most convenient if the test developer can reference
static constants of the actual Android app in her test scripts, especially the
unique identifier constants generated for the Android app (e.g., 
$\texttt{Click}(\texttt{R.id.Btn})$). The second more important reason is Scala's
advance syntactic support for developing domain-specific languages, particularly
implicit values and classes, case classes and object definition, extensible pattern matching,
and support for infix operators. The {\em test controller} module of ChimpCheck
essentially brokers interaction between the test scripts written by the developer
and actual runs of the test on an Android device (hardware or emulator). It is largely
implemented in Scala while leveraging existing and actively maintained libraries:
UI traces of the combinator library are serialized and handed off to Chimp Driver (see below)
via Google's Protobuf library; communications between the test script and
the Android device is managed by the Android testing framework, particularly Adb
and the instrumentation APIs. The Chimp Driver module is the test driver that is
deployed on the Android device. It is implemented as an interpreter that
reads and executes the UI traces expressed by the developer's test script---leveraging on the Android JUnit test runner framework and
the Android Espresso UI exercising framework.
Most importantly, it implements the operational semantics defined in 
Figure~\ref{fig:trace_semantics}. 

\newsavebox{\SBoxEspressoOnView}
\begin{lrbox}{\SBoxEspressoOnView}\small
\begin{lstlisting}[language=Java]
Espresso.onView(
  ViewMatchers.withText("Button1")
).perform(click());
\end{lstlisting}
\end{lrbox}

\myparagraph{Running UI Events in Tandem with the Test App}

As we mention in Section~\ref{ssec:trace-sem}, for reliability of test scripts, 
the test driver runs in tandem with the test app. To account for this design,
our operational semantics imposes the $\idleOrac{\devst}$ restriction on sequencing transitions
(rule $\rname{(Seq-2)}$ that applies to $\trstop \aseq \tr$).
Our implemenation realizes this semantics largely thanks
to the design of the Espresso testing framework: primitive and concrete actions
of our combinator library ultimately map to Espresso actions. For instance, the combinator
$\texttt{Click("Button1")}$ maps to the following Java code fragment that calls the
Espresso library:
\[
\usebox{\SBoxEspressoOnView}
\]
We need not implement any additional synchronization routines
because within the call to $\texttt{perform}$, Espresso
embeds a sub-routine that blocks the control flow of the tester program until the test apps 
main UI thread is in an idle state. This work is essentially the implementation of the $\idleOrac{\devst}$ condition in our
operational semantics. Similarly, property assertions of our operational semantics are required to
be exercised at the appropriate times, and this effect is realized by mapping our property assertions
(i.e., rules $\text{(Assert-Pass)}$, $\text{(Assert-Fail)}$, $\text{(Qualified)}$ and $\text{(Unqualified)}$
for combinators $\texttt{assert}~\prop$ and $\prop~\texttt{then}~\tr$) to Espresso assertions via its 
$\texttt{ViewAssertion}$ library class.

Handling non-concrete events (i.e., events with the UI element wild card \texttt{*}as arguments), however,
requires some more care. To infer relevant UI elements, we need to access the app's run-time view hierarchy.
In order to access the view hierarchy at the appropriate, current state of the app,
our access to the view hierarchy must be guarded by a synchronization routine similar to the ones provided by the Android Espresso library.

\myparagraph{Inferring Relevant UI Elements from the View Hierarchy}

We describe how we realize the rules $\text{(Inferred)}$ and $\text{(Block-2)}$ of the
operational semantics.
As noted above, implementing the inference property of the \texttt{*} combinator relies on accessing the test app's
current view hierarchy. The view hierarchy is a run-time data structure that represents the current
UI layout of the Android app. With it, we can enumerate most UI elements that are currently present
in the app and filter down to the elements that are relevant to the current UI event combinator. For
instance, for $\texttt{Click}(*)$, we want $\texttt{*}$ to be substituted with
some UI element $id$ that is displayed and is clickable, while for $\texttt{Type}(*,s)$, we want
a UI element $id$ that is displayed and accepts user inputs. This filtering is done by leveraging
Espresso's \texttt{ViewMatchers} framework, which provides us a modular way of matching for relevant
views based on our required specifications. Our current implementation is sound in the sense that it will
infer valid UI elements for each specific action type. However, it is not complete: certain UI
elements may not be inferred, particularly because they do not belong in the test app's view
hierarchy. For instance, UI elements generated by the Android framework's dialog library
(e.g., \texttt{AlertDialog}) will not appear in an app's view hierarchy. Our current prototype
will enumerate the default UI elements found in these dialog elements, but it does not attempt to
extract UI elements introduced by user-customized dialogs.

\begin{figure*}\centering\small
\(\begin{array}{@{}c@{}}
 \text{(String Generators)}~\gen^\strD \qquad \text{(Integer Generators)}~\gen^\intD \\ \\[-1.5ex]

 \text{(XY Coordinates)}~\gen^\loc ::= (n,n) \mmid (\gen^\intD,\gen^\intD) 
               \quad\quad \text{(UI Identifiers)}~\gen^\vid ::= s \mmid n \mmid \loc \mmid \wild 
                          \mmid \gen^\strD \mmid \gen^\intD \mmid \gen^\loc \\ \\[-1.5ex]
  \text{(App Events)}~\gen^a  
       ~::=~ \texttt{Click}(\gen^\vid) \mmid \texttt{LongClick}(\gen^\vid) \mmid 
             \texttt{Type}(\gen^\vid,\gen^\strD) \mmid \texttt{Swipe}(\gen^\vid,\gen^\loc)
             \mmid \texttt{Pinch}(\gen^\loc,\gen^\loc)
             \mmid \texttt{Sleep}(\gen^\intD) \\ \\[-1.5ex]
  \text{(Trace Generators)}~\gen ~::=~ 
           \gen^a \mmid \texttt{Skip} \mmid d \mmid \gen \aseq \gen' \mmid 
           \gen~\choice~\gen' \mmid 
           \gen\try \mmid
           \prop~\texttt{then}~\gen \mmid \texttt{repeat}~n~\gen \mmid \trstop
\end{array}\)
\caption{Lifting UI traces from Figure~\ref{fig:traces_and_properties} to UI trace generators.}
\label{fig:generators}
\end{figure*}

\myparagraph{Built-in and User-Defined Predicates}

Primitives of the properties of our language, particularly predicates, are implemented in two 
forms: {\em built-in} predicates are first-class predicates supported by the combinator library.
For example, $\texttt{isClickable}(\vid)$ and $\texttt{isDisplayed}(\vid)$ are both combinators
that accept $\vid$ as an argument, and they assert that the UI element corresponding to $\vid$ is clickable
and displayed, respectively. Our current built-in predicate library includes the full range of
tests supported by Espresso library's $\texttt{ViewMatchers}$ class and their implementations
are simply a direct mapping to the corresponding tests in the Espresso library. 

Our library also allows the test developer to define her own predicates 
(e.g., \texttt{mediaPlayerIsPlaying} from Figure~\ref{fig:mediaplayer-eg2}).
The combinator library provides a generic predicate class that the developer can use
to introduce instances of a predicate, such as \[\texttt{Predicate("mediaPlayerIsPlaying")} \]
or extend the framework with her own case classes that represent predicate combinators.
The current implementation supports predicates with any number of call-by-name arguments of
simple types (e.g., integers, strings).
The developer is also required to provide Chimp Driver an operational interpretation of this predicate 
in the form of a boolean test method of the same name. A run-time call to this test operation is 
realized through reflection.
\revise{The use of reflection is admittedly a possible source of errors in test suite itself because it circumvents static checking. 
To partially mitigate this problem,
ChimpCheck's run-time system has been designed to fail fast and to be explicit
about such errors.
Explicit language support like 
imposing explicit declarations of and pre--run-time checks on user-defined predicates could be introduced
to further mitigate this problem, but such engineering improvements are beyond the scope of a research prototype.}

\myparagraph{Resuming the Test App After Suspending}

Device events like $\texttt{ClickBack}$, $\texttt{ClickHome}$, $\texttt{ClickMenu}$ and 
$\texttt{Settings}$ potentially map to actions that suspend the test app. Our operational semantics silently 
assumes that the app is subsequently resumed so that the rest of the UI test sequence can proceed as planned.
However, implementing this behavior in Espresso, as well as Robotium, is currently not possible within the testing
framework. Instead, to overcome this limitation of the underlying testing framework, we embed 
in our test controller a sub-routine that periodically polls the Android device on what app is currently displayed 
on the foreground. If this foreground app is determined to be one other than the test app, this sub-routine simply 
invokes the necessary commands to resume the test app. These polling and resume commands are invoked through Android's 
command-line bridge (Adb) and this ``kick back'' subroutine is kept active until the UI test trace has been completed (with either success or failure).

\mysection{UI Trace Generators}
\label{sec:trace-generators}

\comments{
We discuss trace generators. Generators are arguably the most important
part of the combinator library as they are the developers main 
interface\footnote{A developer may choose to only rely on UI trace combinators from the previous section, however
ChimpCheck is most effective when programming at the level of generators.} to the test scripts
and provides the developer the means of expressing a family of UI traces that will stress
test his/her test app.
}

We now discuss lifting UI traces to {\em UI trace generators}. In 
Section~\ref{ssec:lang-gen}, we
define the language of generators, by (1) lifting from the symbols of UI traces and (2) introducing
new combinators. Then in Section~\ref{ssec:sem-gen}, we present the
semantics of generators in the form of a transformation operation into the domain of sets of
UI traces. Finally in Section~\ref{ssec:generator-implement}, we present the full architecture of ChimpCheck,
connecting generators to actual test runs and to test results.

\mysubsection{A Language of UI Trace Generators}
\label{ssec:lang-gen}

Figure~\ref{fig:generators} introduces the core language of trace generators. In essence, it is built on top of 
the language of traces (Figure~\ref{fig:traces_and_properties}), particularly extending the formal language in two ways: 
(1) extending primitive event arguments with generators and (2) a richer array of combinators, particularly 
non-deterministic choice and repetition. Our usual primitive terms (coordinates $\gen^\loc$ and UI identifiers
$\gen^\vid$) are now extended with string generators $\gen^\strD$ and integer generators $\gen^\intD$. Formally,
we define string and integer generators as any subset of the string and integer domains, respectively. Hence
app events $\gen^a$ can now be associated to sets of primitive values. Trace generators $\gen$ consists of this extension of
app events ($\gen^a$), the device events $d$, combinators lifted from UI traces (i.e., $\aseq$, $\try$,
and $\texttt{then}$), as well as two new combinators; $\gen \choice \gen'$ ing a non-deterministic
choice between $\gen$ and $\gen'$ and $\texttt{repeat}~n~\gen$ representing the repeated sequencing of 
instances of $\gen$ for up to $n$ times. 

Notice that the combinators $\texttt{optional}$ and $\iseq$ shown in the example in Figure~\ref{fig:mediaplayer-eg2} 
are not part of the core language introduced in Figure~\ref{fig:generators}. The reason is that
they are in fact {\em derivable} from combinators of the core language. For instance, $\texttt{optional}~\gen$ can easily be derived as a non-deterministic choice between traces from $\gen$ or $\texttt{Skip}$
(i.e., $\texttt{optional}~\gen ~\defeq~ \gen \choice \texttt{Skip}$).
{\em Derived} combinators are introduced to serve as syntactic sugar to help make trace generators more concise. 
More importantly, for more advance generative behaviors, derived combinators provide the main form of encapsulation and extensibility
of the library. We will introduce more such derived combinators in Section~\ref{sec:chimp-in-action} and demonstrate
their utility in realistic scenarios.

\mysubsection{A Semantics of UI Trace Generators} \label{ssec:sem-gen}

\begin{figure}\centering\small
\(\begin{array}{@{}c@{}}
\begin{array}{@{}rcl@{}}
   \gdom{\texttt{Click}(\gen^\vid)} & \defeq & 
   \{ \texttt{Click}(id)\mid id \in \gdom{\gen^\vid} \} \\
   \gdom{\texttt{LongClick}(\gen^\vid)} & \defeq & \{ \texttt{LongClick}(id)\mid id \in \gdom{\gen^\vid} \} \\\\[-2.0ex]
   \gdom{\texttt{Type}(\gen^\vid,\gen^\strD)} &  \defeq & 
     \left\{ \texttt{Type}(id,s) ~
            \begin{array}{|c} 
              id \in \gdom{\gen^\vid} ~ \wedge \\
              s \in \gdom{\gen^\strD}
            \end{array} \right\} \\\\[-2.0ex]
   \gdom{\texttt{Swipe}(\gen^\vid,\gen^\loc)} &  \defeq & 
     \left\{ \texttt{Swipe}(id,l) ~
            \begin{array}{|c}
              id \in \gdom{\gen^\vid} ~ \wedge \\
              l \in \gdom{\gen^\loc} 
            \end{array} \right\} \\\\[-2.0ex]
   \gdom{\texttt{Pinch}(\gen^\loc_1,\gen^\loc_2)} & \defeq &
     \left\{ \texttt{Pinch}(l_1,l_2) ~
            \begin{array}{|c}
              l_1 \in \gdom{\gen^\loc_1} ~ \wedge \\ 
              l_2 \in \gdom{\gen^\loc_2}
            \end{array} \right\} \\\\[-2.0ex]
   \gdom{\texttt{Sleep}(\gen^\intD)} & \defeq & \{ \texttt{Sleep}(n)\mid n \in \gdom{\gen^\intD} \}
   \\
   \gdom{\texttt{Skip}} & \defeq & \{\texttt{Skip}\}
   \\
   \gdom{d} & \defeq & \{d\}  
   \\
   \gdom{\gen_1 \aseq \gen_2} & \defeq & 
    \left\{ \tr_1 \aseq \tr_2 ~
           \begin{array}{|c}
             \tr_1 \in \gdom{\gen_1} ~ \wedge \\
             \tr_2 \in \gdom{\gen_2}
           \end{array} \right\} \\\\[-2.0ex]
   \gdom{\gen\try} & \defeq & \{ \tr\try \mmid \tr \in \gdom{\gen} \} \\
   \gdom{\prop~\texttt{then}~\gen} & \defeq & \{ \prop~\texttt{then}~\tr \mmid \tr \in \gdom{\gen} \} \\
   \gdom{\gen_1 \choice \gen_2} & \defeq & \gdom{\gen_1} \cup \gdom{\gen_2} \\\\[-2.0ex]
   \gdom{\texttt{repeat}~n~\gen} & \defeq &
    \left\{ \bigaseq_{i=1}^{m} \tr_i ~
            \begin{array}{|c}
              \tr_i \in \gdom{\gen} ~\wedge \\
              0 < m \leq n
            \end{array} \right\} 
 \end{array}
\end{array}\)
\caption{Semantics of UI trace generators. Generators are interpreted as sets of UI traces that they generate.}
\label{fig:generators-sem}
\end{figure}

The semantics of generators are defined by $\gdom{\gen}$: given generator
$\gen$, the semantic function $\gdom{\gen}$ yields the (possibly infinite) set of UI traces where each
trace $\tr$ is in the {\em concretization}~\cite{DBLP:conf/popl/CousotC77} of $\gen$ (i.e., is an instance of $\gen$). 
Figure~\ref{fig:generators-sem} defines the semantics of UI trace generators, particularly the definition of 
$\gdom{\gen}$. For app events $\gen^a$ (e.g., $\texttt{Click}(\gen^\vid)$, $\texttt{Type}(\gen^\vid,\gen^\strD)$),
the semantic function $\gdom{\gen^a}$ simply defines the set of app event instances with arguments drawn from the respective
primitive generator domains (e.g., $\gen^\vid$, $\gen^\strD$). For $\texttt{Skip}$ and device events $d$,
$\gdom{\texttt{Skip}}$ and $\gdom{d}$ are simply singleton sets. Similar to primitive app events, 
generators of the combinators $\aseq$, $\try$ and $\texttt{then}$ define the sets of their respective
combinators. For non-deterministic choice $\gen_1 \choice \gen_2$, $\gdom{\gen_1 \choice \gen_2}$ defines
the union of $\gdom{\gen_1}$ and $\gdom{\gen_2}$. Intuitively, this means that its concrete traces can be
drawn from either from concrete traces in $\gen_1$ or $\gen_2$. Finally, $\gdom{\texttt{repeat}~n~\gen}$ 
defines the set containing trace sequences of $\tr_i$ of length $m$, where $m$ is between zero and $n$ and
each $\tr_i$ are concrete instances of $\gen$ (denoted by $\bigaseq_{i=1}^{m} \tr_i$). 

\myparagraph{A Trace Combinator? Or A Generator?}

One reasonable question that likely arises by now is the following, ``Why do the combinators for non-deterministic
choice ($\choice$) and repetition ($\texttt{repeat}$) not have counterparts in the language of UI trace 
(Figure~\ref{fig:traces_and_properties})?'' No doubt for uniformity, it appears very tempting to instead define
all the combinators in the language of UI traces and then the simply lifting all constructs of the language into
generators (as done for $\aseq$, $\try$ and $\texttt{then}$ and app events). Doing so however would
introduce some amount of semantic redundancy. For instance, having $\tr_1 \choice \tr_2$ as a UI trace 
combinator would introduce the following two rules in our operational semantics of Chimp Driver 
(Figure~\ref{fig:trace_semantics}):
\begin{center}\small
\(\begin{array}{@{}c@{}}
 \Rule[(Left)]
    {}
    {\ttransstep{\tmenvE{\tr_1 \choice \tr_2}{\devst}}{\trstop}{\tmenvE{\tr_1}{\devst}} } \quad
 \Rule[(Right)]
    {}
    {\ttransstep{\tmenvE{\tr_1 \choice \tr_2}{\devst}}{\trstop}{\tmenvE{\tr_2}{\devst}} }
\end{array}\)
\end{center}
While these rules do offer a richer ``dynamic'' (from the perspective of the single-trace semantics) form of non-de\-ter\-min\-is\-tic 
choice, they provide no additional expressivity to the overall testing framework since the non-deterministic choice
behavior is already modeled by the generator semantics (Figure~\ref{fig:generators-sem}). Similarly, the
repetition ($\texttt{repeat}$) generator faces the same semantic redundancy if it is introduced as a trace combinator.
But conversely, are the UI trace combinators (i.e., $u$, $\aseq$, $\trstop$, $\texttt{assert}$, $\try$ and $\texttt{then}$) 
that we have introduced truly necessarily part of that language?
In fact, they each are indeed necessary: primitive UI events $u$ directly interacts with the device state $\devst$ hence
must inevitably be part of the UI trace language. We need a fundamental way to express sequences of events, hence we have
$\tr_1 \aseq \tr_2$ with $\trstop$ as the unit operator. Finally, notice that the remaining combinators each
in its own way explicitly interacts with the device state: $\texttt{assert}~\prop$ conducts a run-time test $\prop$ 
on the device, what prefix of $\tr\try$ to be executed can only be determined at run-time by inspecting the device state $\devst$, and
deciding to execute $\tr$ in $\prop~\texttt{then}~\tr$ depends on the run-time test $\prop$. In essence,
symbols of the UI trace language should necessarily require run-time interpretation from the device or app state,
while the language of generators may contain combinators that can be ``statically compiled'' away.
This strategy of language and semantics minimization between traces and their generators would continue to make sense as ChimpCheck gets extended.

\newcommand{\devicediag}{
   \begin{tikzpicture}[scale=0.9, every node/.style={scale=0.9}]

   \node[draw=black,style={scale=1}] (Coor) at (0,0) { 
   \renewcommand{\arraystretch}{0.8}
   \begin{tabular}{c}
     Test Controller \\
     $\left( 
      \begin{array}{c}
       \text{Scala},~\text{Google Protobuf}, \\
       \text{Android Adb}, \\
       \text{Instrumentation APIs}
      \end{array}
     \right)$
   \end{tabular}
   };

   \node[draw=black,style={scale=1}] (Drv) at (0,-3) { 
   \renewcommand{\arraystretch}{0.8}
   \begin{tabular}{c}
     Chimp Driver \\
     $\left( 
      \begin{array}{c}
       \text{Java},~\text{Android JUnit}, \\
       \text{Android Espresso}
      \end{array}
     \right)$ \\
     $\ttransstar{\tmenvE{\tr}{\devst}}{\trout}{\tmenvE{\tres}{\devst'}}$
   \end{tabular}
   };

   \draw[->,line width=1.2pt] (-0.9,-0.8)--(-0.9,-2.2) node[midway,left]{
   \begin{tabular}{c}
     Protobuf \\
     encoded \\
     $\tr$
   \end{tabular}
   };
 
   \draw[->,line width=1.2pt] (0.9,-2.2)--(0.9,-0.8) node[midway,right]{
   \begin{tabular}{c}
     Protobuf \\
     encoded \\
     $\trout$ and $\tres$
   \end{tabular}
   };

   \end{tikzpicture}
}

\begin{figure}\centering\small
\begin{tikzpicture}[scale=0.9, every node/.style={scale=0.9}]

\node[draw=black,style={scale=1}] (Dev) at (3,2) { 
   \renewcommand{\arraystretch}{0.8}
   \begin{tabular}{c}
     Test \\ Developer 
   \end{tabular}
};

\node[draw=black,style={scale=1}] (Cmb) at (-0.5,0.4) { 
   \renewcommand{\arraystretch}{0.8}
   \begin{tabular}{c}
     Combinator and \\
     Generator
     Library \\
     (Scala, ScalaCheck)
   \end{tabular}
};

\node[draw=black,style={scale=1}] (Mis) at (5,0.4) { 
   \renewcommand{\arraystretch}{0.8}
   \begin{tabular}{c}
     Coordinator Library \\
     (Scala, Akka Actors)
   \end{tabular}
};

\node[draw=black,style={scale=1},fill=white] (Mut) at (1,-3.9) { \devicediag };
\node[draw=black,style={scale=1},fill=white] (Shd1) at (1.2,-4.1) { \devicediag };
\node[draw=black,style={scale=1},fill=white] (Shd2) at (1.4,-4.3) { \devicediag };

\draw[->,line width=1.2pt] (Dev.west) -| (Cmb.north) node[midway,above] 
     {Develops test generators} ;
\draw[->,line width=1.2pt] (Mis.north) |- (Dev.east) node[midway,right] {
   \begin{tabular}{c}
     Triages \\
     outcomes \\
     $\forall i~\{ \trout_i ~\text{and}~ \tres_i \}$
   \end{tabular}
} ;
\draw[->,line width=1.2pt] (Cmb.east)--(Mis.west) node[midway,above]{
   \begin{tabular}{c}
     UI trace \\
     samples $\tr_i$
   \end{tabular}
};

\draw[->,line width=1.2pt] (Shd2.east)-|(5,-0.1) node[midway,right] {
   \begin{tabular}{c}
     Report \\
     $\trout_i$ and $\tres_i$
   \end{tabular}
}; 

\draw[-,line width=1.2pt] (4,-0.1)|-(2,-1.1);
\draw[->,line width=1.2pt] (2,-1.1)-|(Mut.north) node[midway,above] {
   \begin{tabular}{c}
     Schedule test 
     run $\tr_i$
   \end{tabular}
};

\end{tikzpicture}
\caption{Architecture of ChimpCheck test generation. The test developer
focuses on writing ChimpCheck UI trace generator scripts. At run time, the
underlying ScalaCheck library samples UI traces from these UI trace generators.
The test coordinator issues each UI trace sample to a unique Android emulator, 
which independently exercises an instance of the test app as dictated by the UI trace. The outcome of executing each UI trace is, in the end, reported back to the test developer.}
\label{fig:archit-2}
\end{figure}
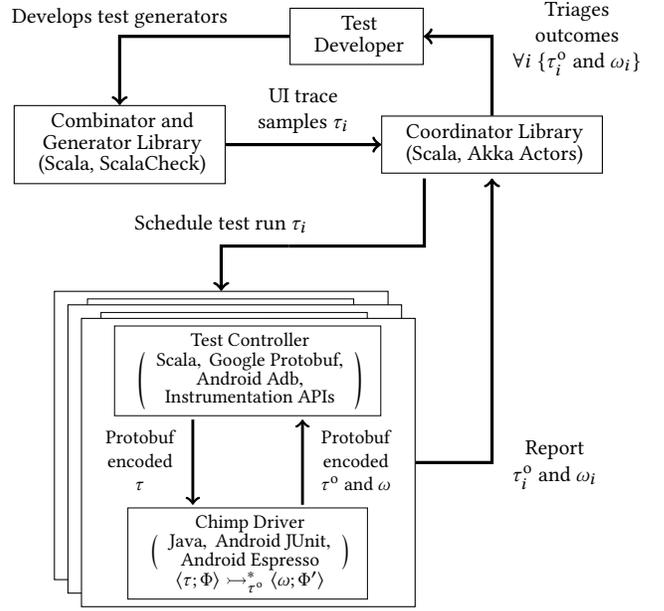

\mysubsection{Implementing UI Trace Generators}
\label{ssec:generator-implement}

Figure~\ref{fig:archit-2} illustrates the entire test generation and
execution architecture of ChimpCheck. The test developer now writes UI trace generators rather than
a single UI trace (as natively in Espresso or Robotium). To generate concrete instances of traces from these generators, we have implemented
the generators on top of the ScalaCheck~\cite{scalacheck} library---Scala's active implementation of
property-based test generation~\cite{DBLP:conf/icfp/ClaessenH00}. Since the testing framework now potentially 
needs to manage multiple test executions, the ChimpCheck
library includes a coordinator library (called Mission Control) that coordinates the test executions:
it coordinates the testing efforts across multiple instances of the test controller and Android devices 
(physical or emulated), scheduling test runs of traces $\tr_i$ across the test execution pool,
and consolidating concretely executed trace $\trout_i$ and test result $\tres_i$ from executing trace $\tr_i$. This coordinator library
is developed with Scala's high-performance Akka actor library.

\myparagraph{Initializing Device State}

Our testing framework architecture assumes that every run of the new test
$\ttransstar{\tmenvE{\tr}{\devst}}{\trout}{\tmenvE{\tres}{\devst'}}$,
is done starting from some initial app state $\devst$. To achieve this,
the default schedule loop of the coordinator library conservatively re-installs the
test app onto the device instance on which the test is to be executed. In general, this re-installation is
the only way to fully guarantee that the test app has been started from a fresh state
However, this default re-install routine can be omitted at
the test developer's discretion. 
\revise{%
Similarly, the developer can specify if devices (for emulators only) should be
re-initialized before starting each test run. Regardless of these initialization
choices, the developer is expected to treat each test run in isolation, and a
single UI trace should be defined to be self-contained, for instance,
including the encoding of time delays for instrumenting the invocation of a 
specific time-idle based behavior in the app. 

Apart from these global configurations, explicit support to control these 
initialization conditions are out-of-scope of the current prototype, though
we postulate possible future work that involve extending the language and
combinator library to allow the developer to specify these initialization conditions
as first-class expressions. This would allow the developer to define more concise
initialization sequences to stress test her app's ability to handling unfavorable
app start-up conditions.
}

\myparagraph{UI Trace Generators with ScalaCheck}

The ChimpCheck Generator library is implemented on top of the ScalaCheck Library, 
an implementation of property-based test generation, QuickCheck~\cite{DBLP:conf/icfp/ClaessenH00}. 
This design choice has proven to be extremely beneficial for our current and anticipated
future development efforts of ChimpCheck: rather than developing randomized test
generation from scratch, we leverage on ScalaCheck's extensive library support for
generating primitive types (e.g., strings and integers denoted by $\gen^\strD$ and 
$\gen^\intD$, respectively). This library support includes many utility functions from 
generating arbitrary values of the respective domains, to uniform/weighted choosing 
operations. The ScalaCheck library is also highly extensible, allowing developers to
extend these functionalities to user-defined algebraic datatypes
(e.g., trees, heaps) and in our case, extending test generation functionalities to 
UI trace combinators. This approach not only
offers the ChimpCheck developer compatible access to ScalaCheck library of combinators,
it makes ChimpCheck reasonably simple to maintain and incrementally developed.

As highlighted earlier in Figures~\ref{fig:mediaplayer-eg1} and~\ref{fig:mediaplayer-eg2},
to generate and execute UI trace test cases, ChimpCheck relies on the 
ScalaCheck library combinator \codecc{forAll} to sample and instantiate 
concrete UI traces, while our $\texttt{chimpCheck}$ combinator embeds
the sub-routines that coordinates execution of the tests (as illustrated in
Figure~\ref{fig:archit-2}). The following illustrates the general form
of the Scala code the developer would write to achieve this:
\begin{lstlisting}[language=ChimpCheck]
 forAll($\gen$: Gen[EventTrace]) { $\tr_i$: EventTrace =>
   $\tr_i$ chimpCheck { $\prop$: Prop }
 }
\end{lstlisting}
UI traces $\tr_i$ are objects of class $\texttt{EventTrace}$ and are sampled and instantiated by
the \codecc{forAll} combinator. The test operation is
\begin{lstlisting}[language=ChimpCheck]
 $\tr_i$ chimpCheck { $\prop$: Prop }
\end{lstlisting}
where $\prop$ is the app property to
be tested at the app state obtained by exercising the app with $\tr_i$. The $\texttt{chimpCheck}$
combinator implements the entire test routine, formally
\[\ttransstar{\tmenvE{\tr_i \aseq \texttt{assert}~\prop}{\devst}}{\trout_i}{\tmenvE{\tres_i}{\devst'}} \;.\]
In addition to ScalaCheck's standard output on total number of tests passed/failed, the ChimpCheck
libraries generate log streams containing the information for reproducing failures, namely
the concrete executed trace $\trout_i$ and test result $\tres_i$. 

\mysection{Case Studies: Customized Test Patterns} \label{sec:chimp-in-action}

In this section, we demonstrate the utility of UI trace generators as a higher-order
combinator library. We present four novel ways that a generator derived from 
the core language is applied to address a specific issue in Android app testing. For each, we discuss 
real, third-party reported issues in open-source apps that motivate the conception of this generator. 

\comments{
\jedi{
Here we show case studies of ChimpCheck used in real apps, each motivated by
a particular strategy to uncovering various common bugs in real Android apps 
and developing regression tests to guard against them. We discuss the following
strategies:
\begin{itemize}
  \item Illegal state exceptions on app resume
  \item Missing/disappearing views
  \item Randomize UI exercising 
  \item Injecting customized scripting into randomized techniques
\end{itemize}
}
Each will motivate the development and discussion of a new combinator 
(from existing ChimpCheck libraries) and highlight of one or more case study apps 
where bug(s) was found with the strategy. The idea is to motivate the utility of
ChimpCheck in not only finding bugs, simplifying test generation, and its
design as an extensible combinator library.
}

\mysubsection{Exceptions on Resume}
\label{ssec:action-illegalstate}

A common problem in Android apps is the failure to handle suspend and resume operations. 
These failures are most commonly exhibited as app crashes caused by (1) {\em illegal state exceptions}, 
when an app's suspend/resume routines do not make correct assumptions on a subcomponent's
life-cycle state, or (2) as {\em null pointer exceptions},
typically when an app's suspend/resume routines wrongly assumes the availability of certain object
resources that did not survive the suspend operation or was not explicitly restored during the resume
operation. From a software testing perspective, the Android app's test suite should include sufficient
test cases that exercises its suspend/resume routines. As illustrated in our example in 
Section~\ref{sec:overview}, Android apps are stateful event-based systems, so conducting suspend and resume
operations at different points (states) of an Android may result in entirely different outcomes.
Test cases must provide coverage for suspend/resume at crucial points of the test app
(e.g., login pages, while performng long background operations).

\myparagraph{The Interruptible Sequencing Combinator}

To simplify the development of trace generator sequences that tests the app's ability to handle 
interrupt events, we derive a specific combinator similar to sequencing but additionally inserts suspend and resume events in a
{\em non-deterministic} manner.

\begin{figure}\centering\small
\(
\begin{array}{l}
   \gen_1 \iseqn{m} \gen_2 ~\defeq~ \gen_1 \aseq \texttt{repeat}~m~\gen_\text{intr} \aseq \gen_2 
   \hfill\text{where} \\[1ex]
   \gen_\text{intr} ~\defeq~
   \texttt{ClickHome} \choice \texttt{ClickMenu} \choice \texttt{Settings} \choice \texttt{Rotate}
\end{array}
\)
\caption{The Interruptible Sequencing Combinator is defined in terms of the
ChimpCheck core language. It sequences $\gen_1$ and $\gen_2$ but allows
a finite number ($m$) of occurrences of interrupt events ($\gen_\text{intr}$) 
in between.}
\label{fig:inter-seq}
\end{figure}

Figure~\ref{fig:inter-seq} shows how the {\em Interruptible Sequencing Combinator} is derived from
the repetition (\texttt{repeat}) combinator: $\gen_1 \iseqn{m} \gen_2$ is defined as sequencing
where we allow $\texttt{repeat}~m~\gen_\text{intr}$ to be inserted between $\gen_1$ and $\gen_2$. The interrupt generator
$\gen_\text{intr}$ denotes the non-deterministic choice between the various interrupt device events
that triggers the suspending (and resuming) of the app. The combinator takes one parameter $m$, which is the
maximum number of times we allow interrupt events to occur. Our current implementation treats this 
parameter optionally and defaults it to $3$, as we have 
observed that in practice, such app crashes are typically reproducible within one or two consecutive interrupt events. 

\myparagraph{Case Studies}

We have observed numerous numbers of issue tracker cases on open Android GitHub projects that report
failures that are exactly caused by this issue (illegal state exceptions or null pointer exceptions
on app resume). One example is found in Tower\footnote{DroidPlanner. Tower. \url{https://github.com/DroidPlanner/Tower}.}, which is a
popular open-source mobile ground control station app for UAV drones. A past issue\footnote{Fredia Huya-Kouadio. FlightActivity NPE in onResume \#1036. \url{https://github.com/DroidPlanner/Tower/issues/1036}. September 2, 2014.}
of the Tower app documents a null-pointer exception that occurs when the user suspended and resumed the app
from the app's main map interface. This failure is caused 
by a wrong assumption that references to the map UI display fragment remain intact after the suspend/resume
cycle.  

Another example worth noting is the Nextcloud open-source Android app\footnote{Nextcloud. \url{https://github.com/nextcloud/android}.}, which
provides rich and secured mobile access to data (documents, calendars, contacts, etc) stored 
(by paid users) in the company's proprietary cloud data storage service. 
A recent issue\footnote{Andy Scherzinger. FolderPicker - App crashes while rotating device \#448. \url{https://github.com/nextcloud/android/issues/448}. December 13, 2016.} reports a crash of the app during a file selection routine when the
user re-orients the app from portrait to landscape. This crash is the result of a null-pointer exception
on a reference to the file selector UI object ($\texttt{OCFileListFragment}$), caused by the failure of
the app's resume operation to anticipate the destruction of the object after suspension.

We have developed test generators for Nextcloud and reproduced this crash (\href{https://github.com/nextcloud/android/issues/448}{\#448}) with the 
following trace generator (with the login sequence omitted for simplicity):
\begin{lstlisting}[language=ChimpCheck]
 $\langle \text{{\rm\em Login Sequence}} \rangle$ *>>
 LongClick(R.id.linearlayout) *>>
 Click("Move") *>> $\trstop$
\end{lstlisting}

Note that other recent work~\cite{DBLP:conf/issta/AdamsenMM15} has also identified injecting interrupt events as being critical to testing apps.
While our (current) implementation of \codecc{*>>} is less sophisticated than  \citet{DBLP:conf/issta/AdamsenMM15}, the advantage of the ChimpCheck approach is that \codecc{*>>} is simply a derived combinator that can be placed alongside other scripted pieces (e.g., for the login sequence).
The \codecc{*>>} provides a basic demonstration of how more complex 
test generators that address real app problems can be implemented in ChimpCheck.

\mysubsection{Preserving Properties}
\label{ssec:enforce-property}

An app's proper functionality may very frequently depend on its ability to preserve
important properties across certain state transitions that it is subjected to. For instance,
it would be a very visible defect if key navigational buttons of the app vanish 
after user suspended and resumed the app.
While this issue seems very similar
to the previous (i.e., a failure caused by interrupt events), the distinction
we consider here is that the issue does not result in app crashes. Hence, simply testing
against interrupt events (via the $\iseq$ combinator) may not detect any faults.
Since the decision of which UI elements should ``survive'' interrupt events is app specific, 
we cannot to fully-automate testing such property but instead derive customizable
generators that allow the test developer to program such tests more effectively and efficiently.

\myparagraph{The Property-Preservation Combinator}

Rather than writing boiler-plate assertions before and after specific events, 
we derive a generator that expresses the test sequences in a more
general manner.

\begin{figure}\centering\small
\(\begin{array}{l}
  \gen~\texttt{preserves}~\prop ~\defeq~ \texttt{assert}~\prop \aseq \gen \aseq \texttt{assert}~\prop
\end{array}\)
\caption{The Property-Preservation Combinator is defined by asserting a 
property $\prop$ before and after a given generator $\gen$, and it tests if
the property is preserved by UI traces sampled from $\gen$.}
\label{fig:preserve-prop}
\end{figure}

\newsavebox{\SBoxPreservesClickable}
\begin{lrbox}{\SBoxPreservesClickable}\small
\begin{lstlisting}[language=ChimpCheck]
$\gen_\textrm{intr}$ preserves isClickable("Button1")
\end{lstlisting}
\end{lrbox}

Figure~\ref{fig:preserve-prop} defines this generator. The Property-Preservation Combinator
$\gen~\texttt{preserves}~\prop$ asserts the (meta) property that $\prop$ is 
preserved across any trace instance of $\gen$. For example, we can assert that \texttt{"Button1"} remains clickable across interrupt events (i.e., after the app resumes):
\begin{center}\small
\(\begin{array}{@{}l@{}}
   \usebox{\SBoxPreservesClickable}\hfill\text{where} \\[1ex]
   \gen_\text{intr} ~\defeq~
   \texttt{ClickHome} \choice \texttt{ClickMenu} \choice \texttt{Settings} \choice \texttt{Rotate} \;.
\end{array}\)
\end{center}

\myparagraph{Case Studies}

We have observed many instances of this ``vanishing UI element'' scenario described above. Just to name a few popular apps in this list:
Pokemap\footnote{Omkar Moghe. Pokemap. \url{https://github.com/omkarmoghe/Pokemap}.},
a supporting map app for the popular Pok\'emon Go game; c:geo\footnote{
c:geo. \url{https://github.com/cgeo/cgeo}.}, a popular Geocaching app with 1M-5M installs from the Google Play Store as of August 23, 2017; and
Tusky\footnote{Andrew Dawson. Tusky. \url{https://github.com/Vavassor/Tusky}.}, an Android client for
a popular social-media channel Mastodon.
Each app at some point of its active development contained a bug related to our given scenario. 
For Pokemap, issue~\#202\footnote{Andy Cervantes. Flipping Screen Orientation Issue - Pokemon Found Stops Being Displayed \#202. \url{https://github.com/omkarmoghe/Pokemap/issues/202}. July 26, 2016.} 
describes
a text display notifying your success in locating a
Pok\'emon permanently disappears after screen rotation. For CGeo, issue~\#2424\footnote{Ondřej Kunc. Download from map is dismissed by rotate \#2424. \url{https://github.com/cgeo/cgeo/issues/2424}. January 22, 2013.} describes
an opened download progress dialog is wrongly dismissed after screen rotation, leaving the 
user uncertain about the download progress. For Tusky, issue~\#45\footnote{Julien Deswaef. When writing a reply, text disappears if app switches from portait to landscape \#45. \url{https://github.com/Vavassor/Tusky/issues/45}. April 2, 2017.} states that 
replies typed into text inputs are not retained after screen rotation.

We found that we could reproduce issue~\href{https://github.com/Vavassor/Tusky/issues/45}{\#45} in Tusky with the following simple generator:
\begin{lstlisting}[language=ChimpCheck]
 $\ldots$ :>> Type(R.id.edit_area,"Hi") :>>
 Rotate preserves hasText(R.id.edit_area,"Hi")
\end{lstlisting}

\comments{
A somewhat related issue is observed in {\em ButtonBar}, 
an open-source Android app library for deploying customizable navigational bars(https://github.com/roughike/BottomBar). 
An issue
($1^{st}~Sep'2016$ issue \#471) documents disappearance of display texts of the navigation buttons
after multiple clicks on any of the buttons. This specific instance of the combinator that will observe this
is as follows:
\[\hspace{-5mm}
\begin{array}{c}
   \texttt{Click(R.id.home)} \choice ... \choice \texttt{Click(R.id.fav)}~\texttt{preserves} \\
   \texttt{isDisplayed("Home")} \wedge ... \wedge \texttt{isDisplayed("Favorites")}
\end{array}
\]

Indeed, the task of writing test generators to find such bugs specifically seem too contrived. 
Instead, the immediate utility of this approach is in regression testing, to guard against recurrence of
similar problems during active development of these apps. 
}
\comments{
On possible future work that can which can leverage
from this combinator is exploring the possibility of synthesizing instances of this combinator
from higher-level user specifications (e.g., on which UI elements should be persistent on which UI 
activities).
}

\mysubsection{Integrating Randomized UI Testing}
\label{ssec:action-random}

While writing custom test scripts is often necessary for achieving the highest possible test coverage,
black-box techniques like pure randomized UI testing~\cite{android-monkey} and model
learning techniques~\cite{DBLP:conf/sigsoft/MachiryTN13,DBLP:journals/software/AmalfitanoFTTM15} are nonetheless important and an effective
means in practice for providing basic test coverage.
Industry-standard Android testing frameworks (e.g., Espresso and Robotium) provides little (or no) support
for integrating with these test generation techniques, which unfortunately forces the test developer to use the various possible testing approaches in isolation.

\myparagraph{The Monkey Combinators}

To demonstrate how black-box techniques can be integrated into our combinator library, here we
derive two generators \texttt{monkey} and \texttt{relevantMonkey} from existing 
combinators.

\begin{figure}\centering\small
\(\begin{array}{@{}l@{}}
 \texttt{monkey}~n \defeq \texttt{repeat}~n~(\gen_\textrm{Ms} \choice \gen_\textrm{intr} ~ \hfill\text{where} \\[1ex] 
 \begin{array}{lcl}
  \gen_\textrm{Ms} & \defeq & \texttt{Click}(\gen^{XY})\try \choice \texttt{LongClick}(\gen^{XY})\try \\
    & \choice & \texttt{Type}(\gen^{XY},\gen^{\strD})\try \choice \texttt{Swipe}(\gen^{XY},\gen^{XY})\try \\
    & \choice & \texttt{Pinch}(\gen^{XY},\gen^{XY})\try \choice \texttt{Sleep}(\gen^\intD)\try
 \end{array} \\\\
 \texttt{relevantMonkey}~n \defeq \texttt{repeat}~n~(\gen_\textrm{Gs} \choice \gen_\textrm{Intr}) ~ \hfill\text{where} \\[1ex]
 \begin{array}{lcl}
  \gen_\textrm{Gs} & \defeq & \texttt{Click}(\wild)\try \choice \texttt{LongClick}(\wild)\try \\
    & \choice & \texttt{Type}(\wild,\gen^{\strD})\try \choice \texttt{Swipe}(\wild,\gen^{XY})\try \\
    & \choice & \texttt{Pinch}(\gen^{XY},\gen^{XY})\try \choice \texttt{Sleep}(\gen^\intD)\try
 \end{array} \\\\
 \gen_\textrm{intr} \defeq
  \texttt{ClickHome} \choice \texttt{ClickMenu} \choice \texttt{Settings} \choice \texttt{Rotate}
\end{array}\)
\caption{Two implementations of the Monkey Combinator. The first (\texttt{monkey}) randomly applies user events on random XY coordinates on a device, mimicking exactly what the Android UI/Application Exerciser Monkey does. The next (\texttt{relevantMonkey}) is just slightly smarter---applying more relevant actions by accessing the device's run-time view hierarchy and inferring relevant user events.}
\label{fig:monkey-gorilla}
\end{figure}

Figure~\ref{fig:monkey-gorilla} shows two implementations of generators for random 
UI event sequences. The first, called the $\texttt{monkey}$ combinator, is similar to
Android's UI Exerciser Monkey~\cite{android-monkey} in that it generates random UI events applied to random locations
on the device screen. The second combinator, called the $\texttt{relevantMonkey}$,
generates random
but more relevant UI events by relying on ChimpCheck's primitive $\texttt{*}$ combinator to
infer relevant interactions from the app's run-time view hierarchy. Having randomized test
generators like these as combinators provides the developer with a natural
programming interface to integrate these approaches with her own custom scripts. For 
instance, revisiting our example in Section~\ref{sec:overview} (or similarly for the Nextcloud app in 
Section~\ref{ssec:action-illegalstate}), 
getting pass a login page is the hurdle to using a brute-force randomized testing, 
but we can implement the necessary traces to
the media pages by simply the following generator:
\begin{lstlisting}[language=ChimpCheck]
 Click(R.id.enter) :>>
 Type(R.id.username,"test") :>>
 Type(R.id.password,"1234") :>>
 Click(R.id.signin) :>> relevantMonkey 50
\end{lstlisting}
This generator simply applies the relevant monkey combinator (arbitrarily for $50$ steps) after the login sequence---thus generating multiple UI traces that randomly exercises the app functionalities after
applying the login sequence fixture.

\myparagraph{Case Studies}

Our preliminary studies have found that even for simple apps
(e.g., Kistenstapeln\footnote{Fachschaft Informatik. Kistenstapeln-Android. \url{https://github.com/d120/Kistenstapeln-Android}.}, a score tracker for crate stacking game,
and Contraction Timer\footnote{Ian Lake. Contraction Timer. \url{https://github.com/ianhanniballake/ContractionTimer}.}, a timer that tracks and stores contraction data), we require generating
numerous UI event sequences from Android Monkey before we get acceptable
coverage results. Preliminary experiments using the relevant monkey combinator (that accesses the view hierarchy) 
have shown promising results shown in Table~\ref{tbl:monkey-exp}. For
the Kistenstapeln app, the relevant-monkey
combinator witnesses the bug from issue~\#1\footnote{Tobias Neidig. Crash on timer-event on other fragment \#1. \url{https://github.com/d120/Kistenstapeln-Android/issues/1}. March 19, 2015.}) in an order of magnitude less
generated events than the Android UI Exerciser Monkey. Note that Android Monkey failed to witness the bug in under 5,000 events in half of the attempts.

\begin{table}\small
\caption{We applied ChimpCheck's \codecc{relevantMonkey} and the Android UI Exerciser Monkey to try to witness a known issue (\href{https://github.com/d120/Kistenstapeln-Android/issues/1}{\#1}) in \href{https://github.com/d120/Kistenstapeln-Android}{Kistenstapleln-Android}. We ran each exerciser 10 times for up to 5,000 UI events. The average number of steps is taken only over successful attempts in witnessing the bug.}
\begin{tabular*}{\linewidth}{@{\extracolsep{\fill}}lrrrr@{}}\toprule
  UI Exerciser & Attempts & \multicolumn{2}{c}{Witnessed} & Steps to Bug \\ \cmidrule(lr){3-4}
               & (n) & (n) & (frac) & (average n) \\
\midrule
  \codecc{relevantMonkey} & 10 & 10 & 1\phantom{.0}       & 289   \\
  Android Monkey     & 10 & 5 & 0.5       & 3177  \\
\bottomrule
\end{tabular*}
\label{tbl:monkey-exp}
\end{table}

We also note that
these promising preliminary results 
are achieved with a simplistic, light-weight implementation: exactly the code in Figure~\ref{fig:monkey-gorilla}, together 
with less than 200 lines of library code that implements view hierarchy access for the $\texttt{*}$ combinator,
developed within a span of two days, including time to learn the Android Espresso framework.

\comments{
We have also tried the gorilla combinator in the Next Cloud app from Section~\ref{ssec:action-illegalstate},
particularly by replacing the hard-coded sequence $\texttt{LongClick(R.id.linearlayout)} \aseq 
\texttt{Click("Move")} \aseq \trstop$ with $\texttt{gorilla 50}$. In this particular case, the gorilla
combinators ability in enumerating possible click activities from the view hierarchy makes it fairly 
simple to achieve the above UI trace in one of its generated cases.
}

\mysubsection{Injecting Custom Generators}
\label{ssec:action-injection}

In Section~\ref{ssec:action-random}, we demonstrated how random UI testing techniques can be added to ChimpCheck as black-box generators. 
However in practice, many situations require more tightly coupled interactions between the custom scripts and 
the black-box techniques. For instance, many modern Android apps can contain features requiring user authentication 
with her account and such authentication procedures are often requested in an on-demand manner (only when user requests contents that requires two-factor authentication). Such 
dynamic behaviors makes it difficult or impractical to simply hard-code and
prepend a custom login script as we did in the previous sections.
 
\comments{
Another example is when we to inject
hard-coded UI sequences in certain situations, for instance if a standard file selector dialog is opened, 
we want to hard-code the non-deterministic choice between a valid and invalid file selection.
}

\myparagraph{The Gorilla Combinator}

From the \codecc{relevantMonkey} combinator, we refine it into the \texttt{gorilla} combinator with the ability to inject customized scripting logic into 
randomized testing.

\comments{
\begin{array}{lcl}
  \gen^{Is} & = & \texttt{Click}(\wild)\try \choice \texttt{LongClick}(\wild)\try \choice \texttt{Type}(\wild,\gen^{\strD})\try \\
    & \choice & \texttt{Swipe}(\wild,\gen^{XY})\try \choice \texttt{Pinch}(\gen^{XY},\gen^{XY})\try \\
    & \choice & \texttt{Sleep}(\gen^\intD)\try \choice \texttt{ClickHome} \choice \texttt{ClickMenu} \\
    & \choice & \texttt{Settings} \choice \texttt{Rotate} \choice \texttt{ClickBack}
 \end{array}
}

\begin{figure}\centering\small
\[\begin{array}{@{}l@{}}
 \texttt{gorilla}~n~\gen \defeq \texttt{repeat}~n~(\gen \aseq (\gen_\textrm{Ms} \choice \gen_\textrm{intr}) ~ \hfill\textrm{where} \\[1ex]
 \begin{array}{lcl}
  \gen_\textrm{Ms} & \defeq & \texttt{Click}(\gen^{XY})\try \choice \texttt{LongClick}(\gen^{XY})\try \\
    & \choice & \texttt{Type}(\gen^{XY},\gen^{\strD})\try \choice \texttt{Swipe}(\gen^{XY},\gen^{XY})\try \\
    & \choice & \texttt{Pinch}(\gen^{XY},\gen^{XY})\try \choice \texttt{Sleep}(\gen^\intD)\try
 \end{array} \\\\
  \gen_\textrm{intr} \defeq 
  \texttt{ClickHome} \choice \texttt{ClickMenu} \choice \texttt{Settings} \choice \texttt{Rotate}
\end{array}
\vspace{-0mm}
\]
\caption{The Gorilla Combinator enriches the monkey combinators with
an additional generator parameter $\gen$. This generator is
injected before every randomly sampled traces from $\gen_\textrm{Ms}$ or
$\gen_\textrm{intr}$.}
\label{fig:custom-gorilla}
\end{figure}

Figure~\ref{fig:custom-gorilla} shows the implementation of the \texttt{gorilla} combinator. 
It accepts an additional argument: a generator $\gen$ that is prepended before every randomized
event $(\gen_\textrm{Ms} \choice \gen_\textrm{intr}$, hence allowing the developer to ``inject'' 
custom {\em directives} to handle situations where a purely randomized technique may be
mostly ineffective. 
For example, we can define a simple combinator that generates traces that randomly explores the app unless a login
page is displayed. When the login page is displayed, it will append a hard-coded login sequence to effectively proceed 
through the login page:
\begin{lstlisting}[language=ChimpCheck]
 val login = Click("Login") :>>
  Type("User","test") :>>
  Type("Password","1234") :>> Click("Sign-in")
 gorilla 50 (isDisplayed("Login") then login)
\end{lstlisting}

%
%

\myparagraph{Case Studies}

\comments{
{\em PwdStore} a password
storage app \url{https://github.com/zeapo/Android-Password-Store}. When we use randomized UI techniques, 
it is likely that we occasionally generate events that sign-outs from the app and having the 
\texttt{subservientGorilla} with a hard-coded login directive (as seen in Figure~\ref{fig:custom-gorilla})
allows for the possibility of generating login and logout loops. 
In another example,}

An example of a simple app with an authentication page is OppiaMobile\footnote{Digital Campus. OppiaMobile Learning. \url{https://github.com/DigitalCampus/oppia-mobile-android}.}, a mobile learning app. 
We have developed test generators using the refined \texttt{gorilla} with a hard-coded login directive 
(as described above). In sample runs, we observed that the \texttt{gorilla}
occasionally logs out and logs back in between unauthenticated and authenticated portions of the page. Though no bugs were
found, such log in/out loops are very rarely tested and potentially hides defects.

This app
also contains an information, modal dialog that is unfavorable for randomized testing techniques. In
particular, the only way to navigate through this page is a lengthly scroll to the end of the UI view
and hitting a button labeled \codecc{"Select SD Card"}. Attempts at random exercising often
ends up stuck in this problematic page. To help the \texttt{gorilla} function more effectively, we 
injected a non-deterministic choice between the only two possible actions when this dialog page is 
visible: (1) proceed forward by scrolling down and click the 
button or (2) hit the device back navigation button. In Figure~\ref{fig:oppia-scroll}, we show the few lines that implements this injection of this application-specific way of getting past a particular modal dialog.

\begin{figure}\small
\begin{lstlisting}[language=ChimpCheck,style=number]
gorilla 100 {
 isDisplayed("SD Card Access Framework") then {
  { Swipe(R.id.scroll,Down) :>>
    Click("Select SD Card")     } <+> ClickBack
 }
}
\end{lstlisting}
\caption{Getting past a modal dialog in the \href{https://github.com/DigitalCampus/oppia-mobile-android}{OppiaMobile} app. This ChimpCheck generator says to do random UI exercising unless the app is showing the \code{"SD Card Access Framework"} page. In this case, inject the specific action to either scroll to the bottom of the page and click a specific button to continue or click the back button.}
\label{fig:oppia-scroll}
\end{figure}

%

Finally, another application of \codecc{gorilla} is that
from Android 6.0 (API level 23) onwards, device permissions (e.g., SD card usage, camera,
location) are granted at run-time, as opposed to on installation. The app developer
is free to decide how these {\em dynamic} permissions are requested. Typically, an app will use a modal dialog box with an acknowledgment and reject button.
Dealing with these dynamic permissions is
straight-forward with the \texttt{gorilla} combinator using code, for example, similar to Figure~\ref{fig:oppia-scroll}.

\mysection{Related Work}

Research in test-case generation for Android apps has largely focused on developing
techniques and algorithms to automate testing while 
providing better code coverage than the industrial baseline, Android Monkey~\cite{android-monkey}.
Evodroid~\cite{DBLP:conf/sigsoft/MahmoodMM14} 
explores the adaption of {\em evolutionary testing} methods~\cite{DBLP:conf/icst/BaresiLM10} 
to Android apps.
MobiGuitar~\cite{DBLP:journals/software/AmalfitanoFTTM15} is a testing tool-chain that automatically
reverse-engineers state-machine navigation models from Android apps to generate test
cases from these models. It leverages a model-learning technique for Android apps
called Android Ripper~\cite{DBLP:conf/kbse/AmalfitanoFTCM12}, which uses a stateful enhancement of 
{\em event-flow graphs}~\cite{DBLP:journals/stvr/Memon07} to handle stateful Android apps.
Dynodroid~\cite{DBLP:conf/sigsoft/MachiryTN13} also develops a model-based approach to generate event sequences.
Similar to our approach
in inferring relevant UI events, Dynodroid uses the app's view hierarchy to observe relevant
actions.
\revise{%
Sapienz~\cite{DBLP:conf/issta/MaoHJ16} uses a multi-objective search-based technique
to generate test cases with the aim of maximizing/minimizing several objective functions key to Android
testing.
Techniques described in this work has been successfully applied in an industrial strength tool, Majicke~\cite{majicke}.}

\revise{%
The works mentioned above each offer a steady advancement of automatic test-case generation.
Given this focus of what's automatable, a perhaps unwitting result has been much less attention on the issue of {\em programmability} and {\em customizability} that we identify in this paper.
To use these automatic test-case generation approaches, a test developer often needs to work around the app-specific concerns in awkward ways.
For instance, to deal with login screens,
\citet{DBLP:journals/software/AmalfitanoFTTM15,DBLP:conf/kbse/AmalfitanoFTCM12} assumes that the test developer provides a set of ``initial'' states
of the app that are past the login screens, which thus allows the authors to focus only on the automatable aspects of the app.
The test developer would then have to rely on other techniques (presumably 
scripting-based techniques) to exercise the app to these ``initial'' states.
ChimpCheck offers the potential to fuse these automatic test-case generation techniques with scripted, app-specific behavior.
Another example of this perhaps unwitting result can be found in~\citet{DBLP:conf/issta/MaoHJ16}. 
This approach describes a test-case generation strategy that is inspired by genetic mutation.
Though the ``genes'' can in principle be customized for specific apps, the authors chose to
focus their experiments only on a set of genes that are known to be generic across all apps.
Much less attention was given to the programmability or customizability
question to, for example, empower the test developer
to express her own customized genes.
In the end, studies~\cite{DBLP:conf/icse/AmalfitanoAFTKM15,DBLP:conf/sigsoft/MachiryTN13,DBLP:journals/software/AmalfitanoFTTM15} on the saturation and coverage of automatic test-case generation 
for Android apps provide evidence for the ChimpClick claim---the need for human insight and app-specific knowledge to generate not just traces but \emph{relevant} traces.}

\comments{
The authors additionally instrumented the Android SDK in order to obtain information
on registered (called {\em broadcast receivers} in the Android framework) background system tasks,
allow Dynodroid to infer relevant system events as well.
}

The most closely related work to ChimpCheck are a few pieces that consider some aspect of programmability in generating UI traces.
\citet{DBLP:conf/issta/AdamsenMM15}
introduces a methodology that enriches existing test suites (Robotium scripts) by systematically
injecting interrupt events (e.g., rotate, suspend/resume app) into key locations of test scripts to 
introduce what the authors refer to as {\em adverse conditions} that are responsible for many failures
in real apps.
\revise{Comparing to our approach, this work can be viewed as a more sophisticated implementation of
the interruptible sequencing operator \;\codecc{*>>} injected into the existing test suite in a fixed manner (replacing all \codecc{:>>}'s with \codecc{*>>}).
Our approach is complementary in that ChimpCheck provides the opportunity to fuse interruptible sequencing with other test-generation techniques in a user-specified manner, and we might adapt their
approaches for systematic exploration of injected events and failure minimization to improve ChimpCheck's library combinators.} \citet{DBLP:conf/mobisys/Hao0NHG14} introduces a programmable
framework for building dynamic analyses for mobile apps. A key component of this work
is a programmable Java library for developing customized variants of random UI exercisers similar to 
the Android Monkey.
Our work shares similar motivations
but differs in that we provide a richer programmable interface 
(in the form of a combinator library), while their work provides stronger support for injecting code and logic for dynamic analysis.

Our approach
leverages ideas from
Quick\-Check~\cite{DBLP:conf/icfp/ClaessenH00}, a property-based test-generation framework for the functional programming language Haskell.
Our implementation is built on top of the ScalaCheck 
library~\cite{scalacheck}, which Scala's implementation of the QuickCheck test framework.
A novel contribution of ChimpCheck over QuickCheck and ScalaCheck is the reification of user interactions as first-class 
objects (UI traces) so that UI trace generators can be defined and sampled from the ScalaCheck library. 


\mysection{Towards a Generalized Framework for Fusing Scripting and Generation} \label{sec:beyond}

In this section, we discuss our vision of a general framework for creating effective UI tests by fusing scripting and generation. While the development of ChimpCheck has provided the first steps to this end, here we discuss steps that would take this approach of fusing scripting and 
generation to the next level. Specifically, we envision augmentations in two incremental fronts:

\begin{enumerate}
  \item {\em Integrating} scripting with state-of-the-art automated test-generation techniques   
        (Section~\ref{ssec:general-auto-gen}), beyond pure randomized (monkey) exercising.
  \item {\em Reifying} test input domains beyond just user interaction sequences (UI traces) 
        (Section~\ref{ssec:general-inputs})
\end{enumerate}

From lessons learnt from the conception and development ChimpCheck, we identify and distill key
 design principles and challenges that we need to overcome. Particularly, the key 
 challenge for (1) is to formally define the semantics of interaction between the scripts that 
 the developer writes and each specific test generation technique in which we fuse. For (2), the 
 key challenges are developing general means of expressing multiple input domains (not only 
 UI traces) and defining the means in which these input streams interact with one another. 
 Generalizing reified input domains also introduces an opportunity for exploring an 
 augmentation on state-of-the-art test generation techniques: generalizing them for 
 generating not just UI traces but also other input sequences (e.g., GPS location updates, 
 event-based broadcasts from other apps).
 In the following subsections, we will discuss ideas on how these challenges can be addressed.

\mysubsection{Generalizing the Integration with State-of-the-Art Automated Generators} \label{ssec:general-auto-gen}

\begin{figure}\centering
\begin{tabular}{@{}c@{}}
\revise{\bf (I) Coarse-grained Fusion} \\[1ex]
\includegraphics[scale=0.255]{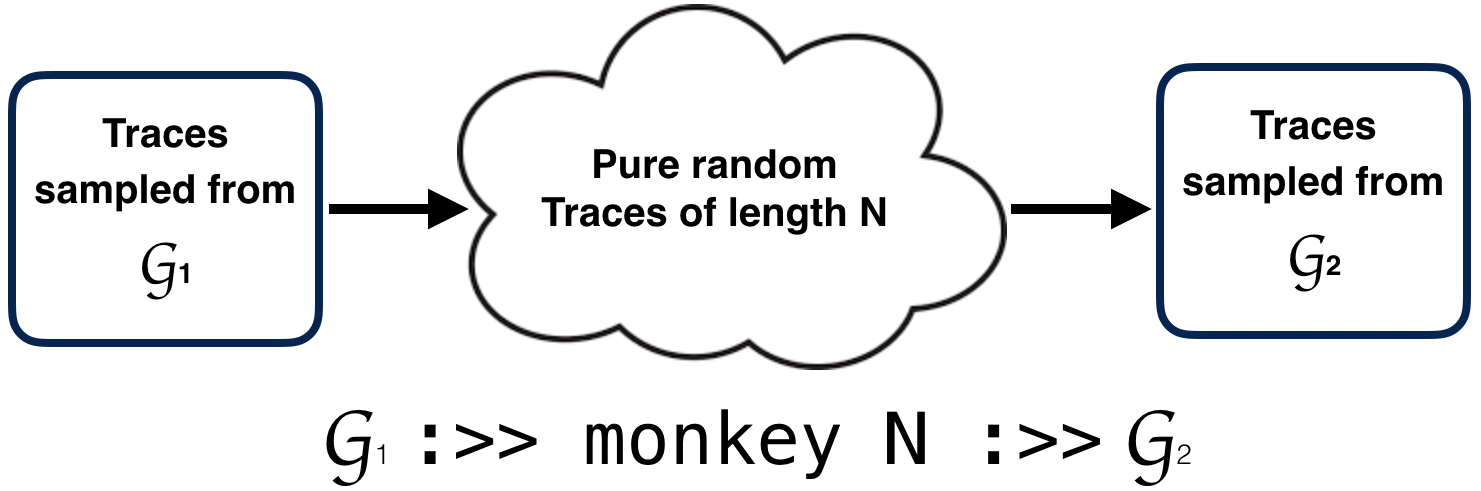} \\ \\[-1ex]
\revise{\bf (II) Fine-grained Fusion} \\[1ex]
\includegraphics[scale=0.255]{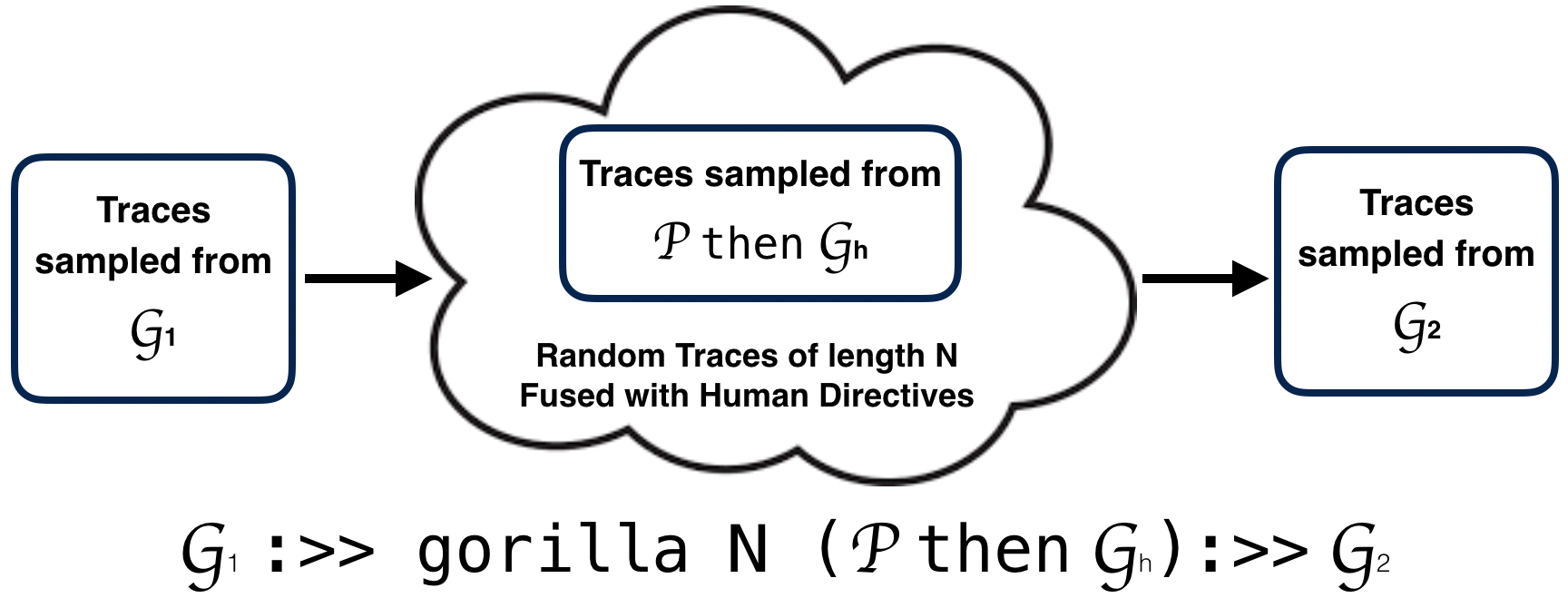}
\end{tabular}
\caption{%
Two ways of using scripting and automated test generation:
(I) the \texttt{monkey} combinator can be invoked within a script, but no 
user interaction is permitted while random exercising is done, 
hence coarse-grained fusion; (II) the \texttt{gorilla} combinator allows 
the test developer to inject human directives into random exercising, hence it is a higher-order combinator enabling finer-grained fusion.}
\label{fig:fuse-monkey-gorilla}	
\end{figure}

We now describe a design principle that will enable us to fuse ChimpCheck
scripts with more advanced forms of automated test generation.
Figure~\ref{fig:fuse-monkey-gorilla} illustrates two ways in which
we have fused scripting and test generation in this paper.
The top most diagram illustrates this interaction implemented by the
\texttt{monkey} combinator (Section~\ref{ssec:action-random}, 
Figure~\ref{fig:monkey-gorilla}). In the diagram,
we represent fragments of the UI traces from scripts
written by the developer ($\gen_1$ and $\gen_2$) with rounded-boxes, 
while the fragment from a randomized exercising technique
($\texttt{monkey~N}$) is represented with a cloud. Edges represent ordering 
between the UI trace fragments, as dictated by the $\aseq$ combinator.
This diagram shows a {\em coarse-grained} interaction between 
user scripts and test generator. Particularly, even though the developer can fuse the
two techniques, the expression $\texttt{monkey~N}$ relinquishes
all control to the randomized exercising technique, other than the parameter 
\texttt{N} that dictates the maximum length of randomized steps.  
In general, this composition can be viewed as a ``uni-directional'' interaction, 
which enables a script to call the automated generator at a specific
point of the UI trace.

\paragraph{Fine-Grained Fusion}
The \texttt{gorilla} (Section~\ref{ssec:action-injection}, 
Figure~\ref{fig:custom-gorilla}) realizes a more {\em fine-grained} interaction: 
shown in the lower diagram of Figure~\ref{fig:fuse-monkey-gorilla}, the \texttt{gorilla} 
combinator refines the \texttt{monkey} combinator by further
allowing the test developer to inject {\em directives} that supersedes
randomized exercising. In this ``bi-directional'' interaction, 
the test developer can interact with the automated test generator by
providing a script that is injected into the randomized exercising
routine. In this instance, the developer injects 
$\prop~\texttt{then}~\gen_h$, customizing the randomized exercising by
stating exceptional conditions ($\prop$) in which a script ($\gen_h$) 
should be used instead of pure randomized exercising. As demonstrated in
Section~\ref{ssec:action-injection}, this combinator enhances
randomized exercising by allowing the developer to guide the test
generation process when needed. The key insight here is that the
higher-order nature of the combinator library (a generator
can be a parameter of another generator $\texttt{gorilla}$) has
enabled us to express this fine-grained interaction that alternates
control between user-directed scripts and the randomized exercising technique.

\begin{figure}\centering
\includegraphics[scale=0.255]{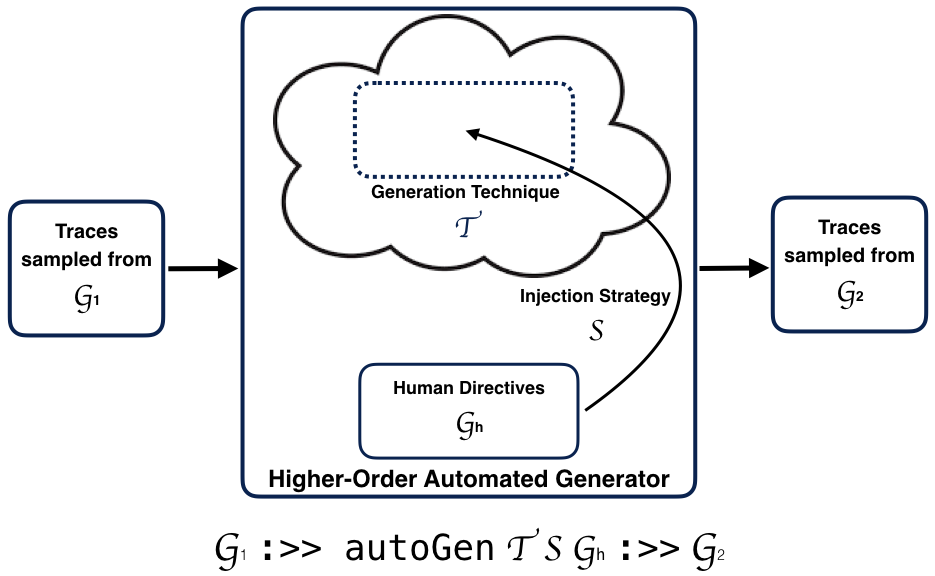}
\caption{\revise{
A higher-order combinator for generalized automated generators.
Parameters ${\mathcal T}$ is an expression that represents an
instance of some automated test-generation technique, while $\gen_h$
is the higher-order generator representing human directives injected
into ${\mathcal T}$. Finally ${\mathcal S}$ specifies the injection 
strategy to be used.
}}
\label{fig:high-order-autogen}
\end{figure}

\paragraph{Generalization}
From a broader perspective, the \texttt{gorilla} combinator is but just
an instance of a higher-order automated generator, specifically
for randomized exercising. We wish to apply this idea to other
state-of-the-art automated test-generation techniques 
(e.g., model-based~\cite{DBLP:journals/software/AmalfitanoFTTM15,DBLP:conf/kbse/AmalfitanoFTCM12},
evolutionary testing~\cite{DBLP:conf/sigsoft/MahmoodMM14}, and 
search-based~\cite{DBLP:conf/issta/MaoHJ16}).
This work would provide a more generalized way for fusing automated test-generation
techniques into the combinator library. Figure~\ref{fig:high-order-autogen} 
illustrates this idea in the form of a combinator $\texttt{autoGen}$.
Similar to the \texttt{gorilla} combinator, it is a 
higher-order combinator that accepts a
generator $\gen_h$. This generator represents the human directives
to be injected into some instance of an automated generator,
which is associated with this call to $\texttt{autoGen}$.
To specify the instance of an automated generator, \texttt{autoGen} takes
in another parameter ${\mathcal T}$, which is a new form of expression
that defines automated test generators. The gorilla combinator can be 
refined as an instance of such an expression (i.e., \texttt{gorilla~N}). 

Parameter ${\mathcal T}$ and $\gen_h$ do not yet completely
define this generalization. Our definition of
\texttt{gorilla~N} earlier in this paper implements a very
specific strategy for injecting $\gen_h$ into 
the randomized exercising technique. Particularly, it treats
the semantics of randomized exercising as a transition system
that appends a new user action (e.g., $\texttt{click}(\vid)$) 
to a UI trace at each derivation step. The strategy of injection
is simply to inject $\gen_h$ between every step. This strategy, we 
call {\em step-wise interleaving},
is effective for fusing with randomized exercising
but not necessarily for other techniques. Hence, in order to
enable more generality, we might need to introduce a third parameter ${\mathcal S}$
that defines the {\em strategy of injection}.

We anticipate that defining ${\mathcal S}$ is the main challenge of
implementing this design principle because it is the key
instrument that defines the semantics of the fusion between 
automated generator and scripting. The applicability of a strategy
${\mathcal S}$ to a generator technique ${\mathcal T}$ would in general
depend on ${\mathcal T}$'s fulfillment of certain properties
(e.g., step-wise interleaving strategy will require ${\mathcal T}$ to
be a well-defined transition system). Developing a set of interfaces
that implements this generalization will be the key engineering
challenge. Our initial observation suggests that 
similar to randomized exercising, model-based techniques 
can exploit the step-wise interleaving strategy, though more investigation 
would be required to ascertain if that would be the most effective strategy. 
Search-based techniques, on the other hand, appear to permit a more sophisticated 
and specialized strategy: we can treat $\gen_h$ as a customized set of
UI traces in which we want the technique's multi-objective search 
algorithm~\cite{DBLP:conf/issta/MaoHJ16} to consider as 
basic UI trace fragments that it uses to generate test sequences.
This essentially allows the developer to interact with the search
algorithm---by fusing her own fragments of user interaction sequences 
(expressed by $\gen_h$) into the search-based test generation technique. 

Such ``higher-orderness'' tempts the obvious question: what happens if we inject an automated
generator into another? It is unclear to us, at this moment, whether such interactions 
are useful or if they should be avoided.
Developing an understanding of the semantics of such 
interactions will be a key challenge to address this question, 
as well as to enable us make the most sensible engineering choices.

\mysubsection{Generalizing the Reification of Test-Input Domains} 
\label{ssec:general-inputs}

\begin{figure}\centering
\includegraphics[scale=0.26]{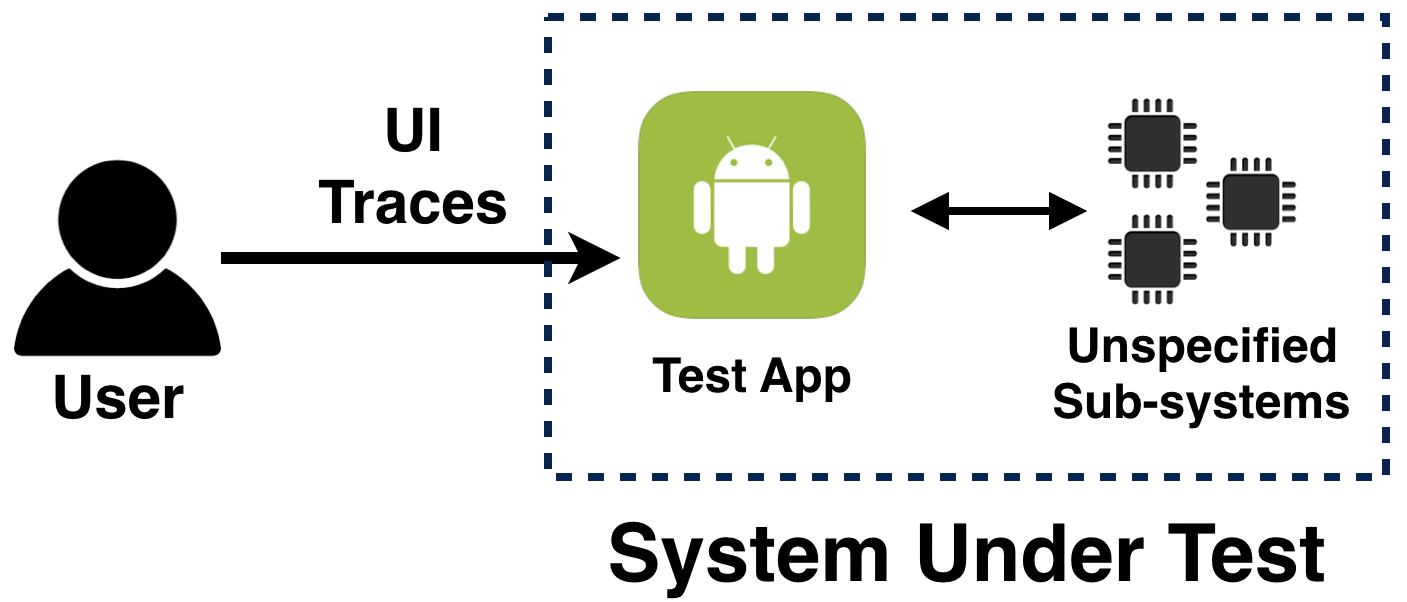}
\caption{\revise{
Reifying user interactions into UI traces---as done in the current ChimpCheck prototype.
}}
\label{fig:chimpcheck-reify}	
\end{figure}

In this section, we discuss fusing 
test scripting and test generation to input domains beyond UI traces 
(user interaction sequences).
To begin, we reflect on the original conception of ChimpCheck's domain-specific language:
since we were interested in user interaction sequences as inputs to our testing efforts, 
we derived symbols that uniquely represent these user actions (e.g., $\texttt{click}(\vid)$).
We next define various combinators (e.g., $\aseq$) that build more complex
UI trace objects from atomic actions.
Now we have means of expressing user interaction sequences (UI traces) as
structured data, which can be manipulated and interpreted. This is the process of 
{\em reification}: concretizing implicit, abstract sequences of events into data structures that 
we can manipulate. For this particular case, we have reified the domain of user-interaction (UI) traces. 
Finally, the language of UI traces is lifted into the language of trace generators,
hence giving us the means of generating and sampling UI traces. 

Figure~\ref{fig:chimpcheck-reify} illustrates an
architectural diagram of the testing strategy for Android apps in 
ChimpCheck. Particularly, the system under test is the test app together
with all other sub-systems that the app interacts with. Since
these other interactions are not reified, test cases are agnostic to their
existence. The edge between the user and test app represents the only input 
into the system under test and is essentially what we have reified into 
UI traces. By reifying UI traces, ChimpCheck 
is able to substitute an actual user with UI traces, and by lifting UI 
traces into the language of UI trace generators, the test developer
has the means of expressing customized UI trace generators for their test apps. 

\begin{figure}\centering
\includegraphics[scale=0.26]{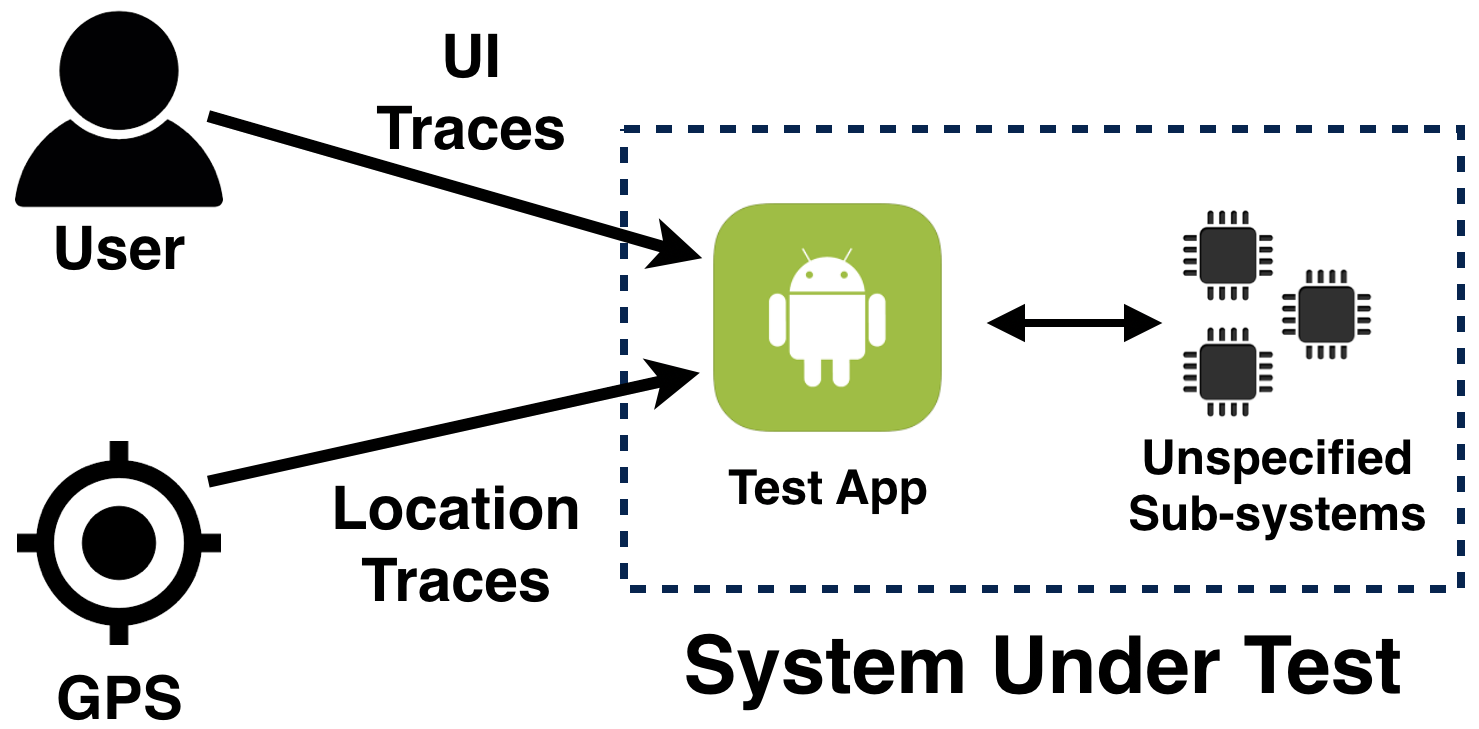}
\caption{\revise{An example of a test environment with two reified domains,
namely UI traces and location traces.}}
\label{fig:custom-reify-eg}
\end{figure}

\paragraph{Reifying Other Domains}
Our key observation is that, while user inputs (UI traces) is arguably 
the most important form of input for user event-driven apps, it is clearly not the only
kind of inputs relevant to testing an app. Reifying other forms of input would enable
us to use similar test-scripting and test-generation techniques to generate input sequences
for testing the app. For instance, by subjecting location updates of the GPS module to the
same reification process, we can
derive the means of generating test cases that simulate the relevant location
updates to the test app---in the same way we have done for UI traces in ChimpCheck.
Figure~\ref{fig:custom-reify-eg} illustrates this new test environment that
has two reified domains: UI traces from the user and location traces from the
GPS module. This generalization introduces a new challenge:
we now have two forms of inputs (UI and location traces), so
how do they interact in terms of the syntax and semantics 
of this generalized language? Our initial observation is that
at earlier stages of testing an app, it is still useful to
derive test cases based on the sequential composition of elements from both domains
(user actions and location updates). Such would be akin to expressing ``laboratory''
tests to test basic functionalities of the app with a controlled (discretized) 
sequence of events. As an example, let us assume the hypothetical reification of 
GPS location updates in the form of a primitive combinator \codecc{locUpdate(lg,la)} where $\texttt{lg}$ and $\texttt{la}$ are simulated inputs (longitude and latitude in decimal
degrees). The following expresses a very specific test sequence in which the the test developer
uses ChimpCheck to inspect the app's handling of a GPS location update after
exercising the app past the login page:
\begin{lstlisting}[language=ChimpCheck]
 Type(R.id.username,"test") :>>
 Type(R.id.password,"1234") :>>
 Click(R.id.login) :>> locUpdate(41.334,2.1567)
\end{lstlisting}
%
 
While the above targets a specific test sequence, it is quite likely that 
at more advanced stages of testing, the test developer might be interested in 
subjecting the app to inputs that the correspond to UI traces interleaved with
location update occurrences, for instance, to test the app's handling of
GPS location updates that interleaves with the user's login sequence.
A parallel composition operator would allow the test developer to express
test generators of this nature, 
exemplified by the following (with a hypothetical parallel compose
operator $\texttt{||}$):
\begin{lstlisting}[language=ChimpCheck]
 { Type(R.id.username,"test") :>>
   Type(R.id.password,"1234") :>>
   Click(R.id.login) } ||
 locUpdate(41.334,2.1567)
\end{lstlisting}
%

Note however, that this parallel composition is necessarily domain restricted.
Other than composing UI traces with location traces,
it might not make sense to allow the developer to parallel-compose streams
of UI traces.
The parallel composition represents an example of an extension of the
language motivated by having multiple input reified domains. In general,
it is likely that more such combinators relevant to and
aimed at capturing idiomatic interactions between various input domains
with be necessary and useful.

\begin{figure}\centering
\includegraphics[scale=0.26]{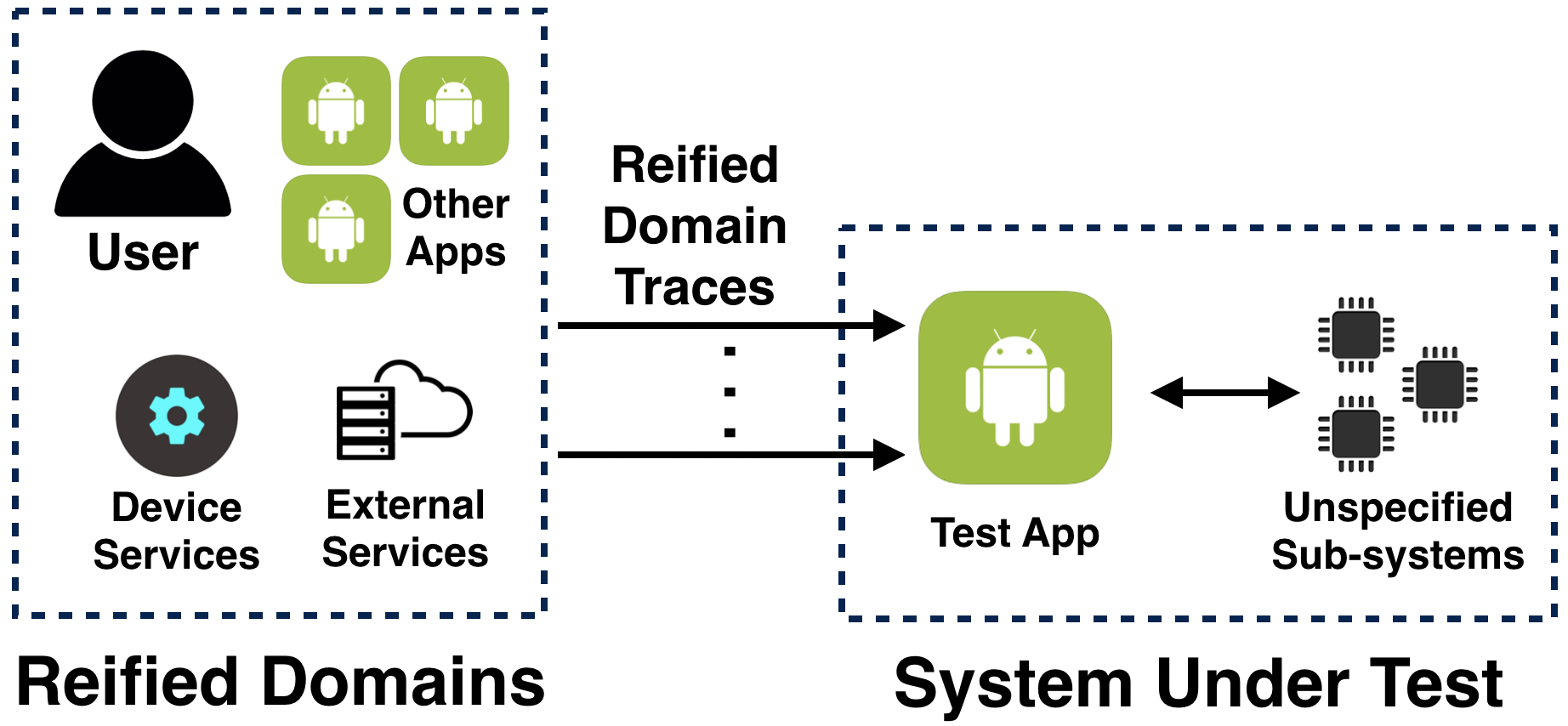}
\caption{\revise{
Generalized test environment where multiple domains (e.g., user input,
inputs from device or external services, inputs from other apps) can be
chosen as reified domains.
}}
\label{fig:custom-reify-general}
\end{figure}

\paragraph{Generalization}
Figure~\ref{fig:custom-reify-general} shows the generalization of this
design principle. Particularly, we should assume that the system under test 
is subjected to inputs from any number of arbitrary reified domains.
The relevant domains for reification are 
the same as the domains of unspecified sub-systems within the system under 
test, namely: (1) user inputs, (2) inputs from device or external services, or 
(3) interactions with other Android apps. Depending on the focus of testing, the test developer
should be allowed to decide which sub-systems fall within the system under
test and which should be explicitly subjected to reification.
This design would allow her to express test generators seamlessly derived from any 
of the reified domains in a single language specification. Other (unreified)
input domains would still interact with the test app as usual but are assumed
not to be the main subjects of the test.
In practice, it is likely that UI traces are typically
given special attention (since Android apps are typically user driven), 
though in theory, UI traces can be uniformly treated as but one of the reified 
domains. That is, we can even entirely omit UI traces if it makes sense for the 
testing needs. An interesting effect of this generalization is that we now
introduce traces of other domains (other than UI traces) to automated
generator techniques, which so far has mostly been studied in the
context of user-interaction sequences. While more studies
are required to understand these new interactions with existing automated test 
generators, we believe that this also constitutes an opportunity to define 
automated generators in a more general manner: allowing us to generate input 
sequences not limited to just UI traces but also compositions of an arbitrary number
of input domains.

We anticipate a number of exciting research opportunities and challenges 
to achieve this dimension of generalization in our testing framework: 
other than extending our combinator library with new reified domains, 
we must facilitate the possibility of allowing the
user to extend the language with her own reified domains.
We expect that a robust library will include reification from common
domains (e.g., user interactions, GPS, WiFi), but the test developer would likely want to 
define, for instance, reified interfaces to her proprietary web service
that her test app calls on during normal usage. This capability will entail
designing programming interfaces that allow the test developer 
to implement the various required runtime obligations of her reified domains.
Another interesting challenge would be to develop ways for the developer to 
express {\em domain-specific constraints} that governs the sampling strategy 
from test generators of each input domain. This ability is important because the notion
 of {\em relevance} of a test input sequence depends heavily on the input 
 domain. For instance, a GPS location update sequence that instantaneously 
 alternates between opposite ends of the Earth is likely to be an unrealistic 
 input sequence. We envision 
 that an effective test-scripting and generation library must include explicit 
 support to enable the test developer to express such domain-specific omissions 
 as constraints over the sampling strategies of the test-generation 
 techniques. Finally, since test inputs are qualified from different domains, it makes sense to introduce explicit
support for asserting certain meta-level properties. For example, a script must adhere to certain
security policies and safety properties when executing inter-app
communication (e.g., a malicious app is present or a dependent app is not installed). Exploring such new
language features will be critical to making the testing framework
expressive and effective in practice.

Implementing the interfaces of new reified domains to testing framework is no doubt often
a tedious endeavor. For instance, in the case of UI traces, we have to develop
a mapping from UI trace atoms to Android Espresso method calls to realize the 
actual exercising on the test app. For the GPS module, we would have to develop
a similar mapping from our reified domain of location updates to
APIs in the \texttt{LocationManager} framework class of the Android framework. 
We believe that by encouraging the development of these mappings as library code that
are accessible by highly
reusable combinators, we ultimately provide more opportunities for reusability.

\mysection{Conclusion}

We considered the problem of exercising interactive apps to generate relevant
user-interaction traces. Our key insight is a lifting of UI scripting from traces to generators
drawing on ideas from property-based test-case generation~\cite{DBLP:conf/icfp/ClaessenH00}. In
particular, we formalized the notion of user-interaction event sequences (or UI
traces) as a first-class object with an operational semantics that is then
implemented by a platform-specific component (Chimp Driver). First-class UI
traces naturally lead to UI trace generators that abstract sets of UI traces.
The sampling from UI trace generators is then platform-independent and can
leverage a property-based testing framework such as ScalaCheck.

Driven by real issues reported in real Android apps, we demonstrated how
ChimpCheck enables easily building customized testing patterns out of
compositional components. The resulting testing patterns like interrupting
sequencing, property preservation, brute-force randomized monkey testing, and a
customizable \texttt{gorilla} tester seem broadly applicable and useful.
Preliminary experiments provide evidence that simple specializations expressed
in ChimpCheck can drive apps to witness bugs with many fewer events.

\revise{
Generalizing from lessons learnt during development and experimentation of ChimpCheck,
we have distilled two design principles, namely higher-order automated generators
and custom reifications of test inputs. We believe that these design principles
will serve as a guide to future development of highly relevant testing
frameworks based on the humble idea of expressing property-based test generators
via combinator libraries.
}

\begin{acks}
We thank the University of Colorado Fixr Team and the Programming Languages and Verification Group (CUPLV) for insightful discussions, as well as the anonymous reviewers for their helpful comments.
This material is based on research sponsored in part by \grantsponsor{GS100000185}{DARPA}{http://dx.doi.org/10.13039/100000185} under agreement number \grantnum{}{FA8750-14-2-0263} and by the \grantsponsor{GS100000001}{National Science Foundation}{http://dx.doi.org/10.13039/100000001} under grant number \grantnum{GS100000001}{CCF-1055066}. The U.S. Government is authorized to reproduce and distribute reprints for Governmental purposes notwithstanding any copyright notation thereon.
\end{acks}

\bibliography{conference.short,bec.short}



\end{document}